\documentclass[a4paper,12pt,twoside]{article}
\usepackage[english]{babel}
\usepackage{rotating}
\usepackage{epsfig}
\usepackage[centertags]{amsmath}
\usepackage{amssymb,amscd,amsatdef}
\usepackage{array}
\usepackage{multirow}
\usepackage{supertabular}
\usepackage{colordvi}
\usepackage{axodraw}
\usepackage{dcolumn}
\usepackage{cite}
\usepackage{xspace}
\usepackage{lscape}
\def\zero{{\scriptscriptstyle 0}}

\def\Z0{\ensuremath{Z^\zero}}

\def\gev{\ensuremath{\,\text{Ge}@!@!@!@!\text{V\/}}}

\def\SU2U1{{\rm SU}(2)\times{\rm U}(1)}

\mathchardef\qsm=63
\mathchardef\pls=43
\mathchardef\mns=512
\mathchardef\plm=518
\mathchardef\eql=61
\mathchardef\smallleft=300
\mathchardef\smallright=301
\mathchardef\perslsh=47
\mathchardef\les=316
\mathchardef\gre=318
\mathchardef\leq=532
\mathchardef\grq=533
\chardef\usc=95
\chardef\til=126

\def\sqr#1#2#3{{\vcenter{\hrule height.#3ex\hbox{\vrule width.#2ex height#1ex
    \kern#1ex\vrule width.#3ex}\hrule height.#2ex}}}

\def\angleto{\vrule width.035em height2.1ex depth-.56ex\unskip\kern-.6ex\to}
\def\perchc#1{{\raise.4ex\hbox{$\mkern4mu#1{\it\perslsh}_
             {\mkern-5mu\scriptscriptstyle{{\rm o}\!{\rm o}}}^
             {\mkern-12.8mu\scriptscriptstyle{\rm o}}$}}}

\catcode`\@=11 
\def\parenbar{\mathpalette\p@renb@r}
\def\p@renb@r#1#2{\vbox{%
  \ifx#1\scriptscriptstyle \dimen@.7em\dimen@ii.2em\else
  \ifx#1\scriptstyle \dimen@.8em\dimen@ii.25em\else
  \dimen@1em\dimen@ii.4em\fi\fi \offinterlineskip
  \ialign{\hfill##\hfill\cr
    \vbox{\hrule width\dimen@ii}\cr
    \noalign{\vskip-.3ex}%
    \hbox to\dimen@{$\mathchar300\hfil\mathchar301$}\cr
    \noalign{\vskip-.3ex}%
    $#1#2$\cr}}}
\catcode`\@=12 

\newbox\struttbox
\setbox\struttbox=\hbox{\vrule height1.65ex depth.485ex width0pt}
\def\strutt{\relax\ifmmode\copy\struttbox\else\unhcopy\struttbox\fi}
\def\stru#1#2{\relax\ifmmode\hbox{\vrule height#1 depth#2 width0pt}
\else\vrule height#1 depth#2 width0pt\fi}

\def\ronum#1{\uppercase\expandafter{\romannumeral#1}}
\def\ronuml#1{\expandafter{\romannumeral#1}}

\DeclareMathAlphabet{\mathbf}{OT1}{cmr}{bx}{sl}

\renewcommand{\baselinestretch}{1.12}

 {\end{list}}
 {\end{list}}
 {\end{list}}

\catcode`\@=11 
\newlength{\@fninsert}
\newlength{\@fnwidth}
\renewcommand{\@makefntext}[1]%
  {\noindent\makebox[\@fninsert][r]{\@makefnmark}\hfil%
  \parbox[t]{\@fnwidth}{#1}}
\catcode`\@=12 
\newlength{\localtextwidth}
\catcode`\@=11 
\newsavebox{\tmpbox}
\newlength{\@captionmargin}
\newlength{\@captionwidth}
\newlength{\@captionitemtextsep}
\renewcommand{\@makecaption}[2]%
  {\def\baselinestretch{0.95}%
   \vspace{10.pt}
   \setlength{\@captionwidth}{\localtextwidth}
   \addtolength{\@captionwidth}{-\@captionmargin}
   \sbox{\tmpbox}{{\bf #1:}{\rm #2}}%
   \ifthenelse{\lengthtest{\wd\tmpbox > \@captionwidth}}%
   {\centerline{\parbox[t]{\@captionwidth}%
   {\tolerance=2000\normalsize%
    {\bf #1:}\hspace{\@captionitemtextsep}{\rm #2}}}}%
   {\centerline{{\bf #1:}\kern1.em{\rm #2}}}}
\catcode`\@=12 
\catcode`\@=11 
\renewcommand\section{\@startsection{section}{1}{\z@}%
                                   {-3.5ex \@plus -1ex \@minus -.2ex}%
                                   {2.3ex \@plus.2ex}%
                                   {\normalfont\Large\bfseries}}
\renewcommand\subsection{\@startsection{subsection}{2}{\z@}%
                                   {-3.25ex\@plus -1ex \@minus -.2ex}%
                                   {1.5ex \@plus .2ex}%
                                   {\normalfont\large\bfseries}}
\renewcommand\subsubsection{\@startsection{subsubsection}{3}{\z@}%
                                   {-3.25ex\@plus -1ex \@minus -.2ex}%
                                   {1.5ex \@plus .2ex}%
                                   {\normalfont\large\bfseries}}
\renewcommand\paragraph{\@startsection{paragraph}{4}{\z@}%
                                   {3.25ex \@plus1ex \@minus.2ex}%
                                   {1.2ex \@plus .2ex}%
                                   {\normalfont\normalsize\bfseries}}
\catcode`\@=12 

\newsavebox{\sesbox}
\newlength{\seslen}

\newcommand{\Et}       {\mbox{$E_{T}$}}
\newcommand{\Pt}       {\mbox{$P_{T}$}}

\def\etjet{E_{T,B}^{{\rm jet}}}
\def\etajet{\eta_B^{{\rm jet}}}
\def\phijet{\phi_B^{{\rm jet}}}

\def\sq2{d\sigma_{\rm tot}/dQ^2}
\def\sdijq2{d\sigma_{2+1}/dQ^2}
\def\rat21{R_{2+1}(Q^2)}
\def\setab2{d\sigma_{2+1}/d\eta_B^{{\rm jet},2}}
\def\Journal#1#2#3#4{{#1} {#2} (#3) #4}

\def\SJN{{Sov. J. Nucl. Phys.}}
\def\SPJ{{Sov. Phys. JETP}}

\def\NIM{{Nucl. Instr. and Meth.} A}
\def\NPB{{Nucl. Phys.} B}
\def\PLB{{Phys. Lett.}  B}
\def\PRL{{Phys. Rev. Lett.}}
\def\PRD{{Phys. Rev.} D}

\def\PRE{{Phys. Rep.}}
\def\ZPC{{Z. Phys.} C}

\def\EPC{{Eur. Phys. J.} C}
\def\CPC{{Comp. Phys. Comm.}}
\def\JPG{{J. Phys.}}
\def\etal{{et al.}}

\newcommand{\draftversion}[1]{\gdef\@draftversion{#1}}

\newcommand{\dab}      {_{\rm \scriptscriptstyle DA}}

\newcommand{\as}       {\mbox{$\alpha_{s}$}}
\newcommand{\asz}      {\mbox{$\alpha_{s}(M_Z)$}}

\newcommand{\qq}       {\mbox{$Q^{2}$}}

\newcommand{\qqda}     {\mbox{$Q^{2}\dab$}}
\newcommand{\xda}      {\mbox{$x\dab$}}
\newcommand{\yda}      {\mbox{$y\dab$}}

\newcommand{\gevv}     {\mbox{${\rm Ge\kern -0.1em V}^{2}$}}
\begin{document}
\selectlanguage{english}
\thispagestyle{empty}
\title{
\bf\Large Measurement of dijet production in neutral\\
\bf\Large current deep inelastic scattering at high ${\bf Q^2}$\\
\bf\Large and determination of ${\bf \alpha_s}$
\vspace{1.5cm}
}

\author{ZEUS Collaboration}
\date{ }
\maketitle

\vfill
\centerline{\bf Abstract}
\vskip4.mm
\centerline{
  \begin{minipage}{15.cm}
    \noindent
Dijet production has been studied in neutral current deep inelastic $e^+ p$
scattering for 470 $ < Q^2 < $ 20000 GeV$^2$ with the ZEUS detector at HERA
using an integrated luminosity of 38.4~pb$^{-1}$. Dijet differential cross
sections are presented in a kinematic region where both theoretical and
experimental uncertainties are small. Next-to-leading-order (NLO) QCD
calculations describe the measured differential cross sections well. A QCD
analysis of the measured dijet fraction as a function of $Q^2$ allows both
a precise determination of $\asz$ and a test of the energy-scale dependence of
the strong coupling constant. A detailed analysis provides an improved 
estimate of the uncertainties of the NLO QCD cross sections arising from the
parton distribution functions of the proton. The value of $\alpha_s(M_{Z})$, as
determined from the QCD fit, is
$\alpha_s(M_{Z}) = 0.1166 \pm 0.0019 \; {\rm (stat.)}
^{+ 0.0024}_{- 0.0033} \; {\rm (exp.)} ^{+ 0.0057}_{- 0.0044} \; {\rm (th.)}$. 
  \end{minipage}
  }
\vfill

\pagestyle{plain}
\thispagestyle{empty}
\newpage
\def\3{\ss}
\pagenumbering{Roman}
\begin{center}
{\Large  The ZEUS Collaboration }
\end{center}
  J.~Breitweg, S.~Chekanov,  M.~Derrick, D.~Krakauer, S.~Magill, 
  B.~Musgrave, A.~Pellegrino, J.~Repond, R.~Stanek, R.~Yoshida\\
 {\it Argonne National Laboratory, Argonne, IL, USA}~$^{p}$
% ...
\newpage                                                                                           
$^{\    1}$ now visiting scientist at DESY \\                                                      
$^{\    2}$ now at Univ. of Salerno and INFN Napoli, Italy \\                                      
% ...

\newpage   
\begin{tabular}[h]{rp{14cm}}                                                                       
$^{a}$ &  supported by the Natural Sciences and Engineering Research
          Council of Canada (NSERC)  \\
$^{b}$ &  supported by the FCAR of Qu\'ebec, Canada  \\
$^{c}$ &  supported by the German Federal Ministry for Education and
          Science, Research and Technology (BMBF), under contract
          numbers 057BN19P, 057FR19P, 057HH19P, 057HH29P, 057SI75I \\
$^{d}$ &  supported by the MINERVA Gesellschaft f\"ur Forschung GmbH, the
          Israel Science Foundation, the U.S.-Israel Binational Science
          Foundation, the Israel Ministry of Science and the Benozyio Center
          for High Energy Physics\\
$^{e}$ &  supported by the German-Israeli Foundation, the Israel Science
          Foundation, the U.S.-Israel Binational Science Foundation, and by
          the Israel Ministry of Science \\
$^{f}$ &  supported by the Italian National Institute for Nuclear Physics
          (INFN) \\
$^{g}$ &  supported by the Japanese Ministry of Education, Science and
          Culture (the Monbusho) and its grants for Scientific Research \\
$^{h}$ &  supported by the Korean Ministry of Education and Korea Science
          and Engineering Foundation  \\
$^{i}$ &  supported by the Netherlands Foundation for Research on
          Matter (FOM) \\
$^{j}$ &  supported by the Polish State Committee for Scientific Research,
          grant No. 111/E-356/SPUB-M/DESY/P-03/DZ 3001/2000,
          620/E-77/SPUB-M/DESY/P-03/DZ 247/2000, and by the German Federal
          Ministry of Education and Science, Research and Technology (BMBF)\\
$^{l}$ &  partially supported by the German Federal Ministry for
          Education and Science, Research and Technology (BMBF)  \\
$^{m}$ &  supported by the Fund for Fundamental Research of Russian Ministry
          for Science and Edu\-cation and by the German Federal Ministry for
          Education and Science, Research and Technology (BMBF) \\
$^{n}$ &  supported by the Spanish Ministry of Education
          and Science through funds provided by CICYT \\
$^{o}$ &  supported by the Particle Physics and
          Astronomy Research Council \\
$^{p}$ &  supported by the US Department of Energy \\
$^{q}$ &  supported by the US National Science Foundation
\end{tabular}
                                                           %
\newpage
\setcounter{page}{1}
\pagenumbering{arabic}
\section{Introduction}
\vspace{-0.3cm}
\label{sec:intro}
\indent
Dijet production in neutral current (NC) $e^+ p$ deep inelastic scattering
(DIS) provides a rich testing ground of the theory of the strong interactions
of quarks and gluons, namely quantum chromodynamics (QCD). At leading order
(LO) in the strong coupling constant, $\alpha_s$,  2+1 jet
production\footnote{Hereafter we denote the proton remnant by ``$+1$''.} in NC
DIS proceeds via the QCD-Compton ($V^* q \rightarrow qg$ with
$V =\gamma,\,Z^0$) and boson-gluon fusion ($V^* g \rightarrow q \overline{q}$)
processes. Thus, the differential cross section for dijet production is
directly sensitive to $\alpha_s$, which is the fundamental parameter of the
theory. Selecting a phase-space region where the perturbative QCD (pQCD)
predictions are least affected by theoretical uncertainties provides a 
compelling test of QCD and permits a precise determination of the strong
coupling constant. 

In this paper, measurements of the differential cross sections for dijet 
production in NC DIS are presented and compared with next-to-leading-order
(NLO) pQCD predictions after correction for hadronisation effects. The
phase-space region is restricted to high values of the virtuality, $Q^2$, of
the exchanged boson, $470 < Q^2 < 20000$~GeV$^2$. In this region, the
experimental uncertainties on the reconstruction of both the positron and the
hadronic final state are smaller than at lower $Q^2$. In addition, the
theoretical uncertainties due to the modelling of the hadronic final state, to
the parton distribution functions (PDFs) of the proton and to the higher-order
contributions are minimised. The comparison of the differential cross sections
with the theoretical predictions of the underlying hard processes provides a
test of the NLO QCD description of dijet production. The analysis takes fully
into account the correlation between the value of $\asz$ used in the
determination of the proton PDFs and that in the calculation of the partonic
cross sections. Furthermore, a detailed study of the theoretical uncertainties
has been carried out which includes the statistical and correlated systematic
uncertainties from each data set used in the determination of the proton PDFs.
The QCD analysis yields a precise determination of $\alpha_s(M_Z)$ and its
energy-scale dependence. The twelve-fold increase in integrated luminosity, in
combination with the improved experimental and theoretical methods used here,
produces a significantly more accurate determination of $\asz$ with respect to
the previous ZEUS measurement \cite{oldalp}.
\vspace{-0.3cm}
\section{Theoretical framework}
\label{sec:qcdpred}
\subsection{Kinematics}
For a given $e^+ p$ centre-of-mass energy, $\sqrt{s}$, the cross section for
NC DIS, $e^+ p \rightarrow e^+ + {\rm X}$, depends on two independent kinematic
variables, which are chosen to be $Q^2$ and the Bjorken scaling variable, $x_{Bj}$, where $Q^2=-q^2$
and $x_{Bj}=Q^2/(2 P \cdot q)$, and $P$ ($q$) is the four-momentum of the
incoming proton (exchanged virtual boson, $V^*$).

For dijet production in NC DIS at LO, three additional variables,
$\xi,z_p$ and $\Phi$, are needed in order to describe the kinematics of the two
outgoing massless partons which form the two jets. The fraction $\xi$ of the proton
four-momentum carried by the incoming parton is defined by
$\xi=x_{Bj} \cdot (1+M^2_{12}/Q^2)$, where $M_{12}$ is the invariant mass of
the two final-state partons. The variable $z_p$ is defined for each outgoing
parton $i$ ($i=1,2$) with four-momentum $p_i$ by
\begin{equation}
z_{p,i} = \frac {P\cdot p_i}{P\cdot q} = 
          \frac {1}{2} \cdot (1-\cos\theta^*_i) \simeq 
          \frac {E_{L}^{{\rm jet},i}(1-\cos\theta_{L}^{{\rm jet},i})}
                {\sum_{k=1,2} E_{L}^{{\rm jet},k}(1-\cos\theta_{L}^{{\rm jet},k})},
\nonumber
\end{equation}
where $\theta^*_i$ is the parton scattering angle in the incoming parton-$V^*$
centre-of-mass system and \mbox{$z_{p,1}+z_{p,2}=1$}. Experimentally,
$z_p$ can be determined from the energy, $E_{L}^{{\rm jet}}$, and polar angle,
$\theta_{L}^{{\rm jet}}$, of each of the two reconstructed jets in the
laboratory frame. The angle $\Phi$ is the azimuthal angle between the parton
and lepton scattering planes in the $V^*$-parton centre-of-mass system.
\subsection{Cross-section calculation}
According to the QCD-improved parton model, as quantitatively expressed by the
factorisation theorem of QCD \cite{factor} and perturbation theory, a NC DIS
differential cross section\footnote{The same symbol, $d\sigma$, is used for both
the DIS inclusive, $d\sigma_{\rm tot}$, and the dijet, $d\sigma_{2+1}$,
differential cross sections. The same symbols with the superscript NLO
indicate only the perturbative component (the first term on the right-hand
side of Eq.~(\ref{eq:fact})).}, $d\sigma$, can be written as
\begin{equation}
  d\sigma=\sum_{a=q,\overline{q},g}\int dx 
  f_{a}(x,\mu_F^{2};\as;\{\zeta\}) \cdot
  d\hat{\sigma}_{a}(x P,\as(\mu_R),\mu_R^{2},\mu_F^{2})\;
    \cdot (1+\delta_{{\rm had}}),
  \label{eq:fact}
\end{equation}
which can be interpreted as follows. The cross section has the form of a
convolution of the partonic hard cross sections, $d\hat{\sigma}_{a}$, with the
PDFs, $f_{a}(x,\mu_F^{2})$, of the colliding proton, with respect to the
fraction $x$ of the proton four-momentum carried by the incoming parton. The
partonic cross sections describe the short-distance structure of the
interaction and are calculable as power-series expansions in the strong
coupling constant, which depends on the renormalisation scale, $\mu_R$. The
PDFs contain the description of the long-distance structure of the incoming
proton. Their evolution with the factorisation scale, $\mu_F$, at which they
are probed follows the DGLAP equations \cite{dglap}. In Eq.~(\ref{eq:fact}), the
dependence of the PDFs on the value of the strong coupling constant assumed in
the DGLAP equations and on the parameters (collectively indicated by 
$\{ \zeta \}$) needed to model the $x$ dependence of the PDFs is explicitly 
indicated. In this analysis, the implicit $\alpha_s(M_Z)$ dependence of the
PDFs in the calculation of the NLO differential cross sections has been taken
into account. The non-perturbative contribution, $\delta_{{\rm had}}$, to the
inclusive DIS cross section, $\sigma_{\rm tot}$, can be safely neglected in
the high-$Q^2$ region studied here. For the differential dijet cross sections,
$\delta_{{\rm had}}$ was estimated using Monte Carlo (MC) models for the
parton cascade and fragmentation (see Section~4).

In the past few years there has been
considerable theoretical progress on the development of general (i.e. process- 
and observable-independent) algorithms that allow a complete analytical
cancellation of the  soft and collinear singularities encountered in the
calculation of NLO jet cross sections \cite{singu,disent}; flexible programs
that compute arbitrary infrared- and collinear-safe DIS observables in NLO QCD
are now available \cite{disent,mepjet,disaster,jetvip}.
\vspace{-0.3cm}
\section{Experimental setup}
The data sample was collected with the ZEUS detector~\cite{status}
during the 1996-1997 data-taking period at HERA and corresponds to
an integrated luminosity of $38.4\pm 0.6$~pb$^{-1}$. During this period HERA
operated with protons of energy $E_p=820$~GeV and positrons of energy
$E_e=27.5$~GeV. 

The compensating uranium-scintillator calorimeter (CAL) \cite{zeuscal} covers
$99.7\%$ of the total solid angle. It is divided into three parts with respect
to the polar angle\footnote{The ZEUS coordinate system is a right-handed
Cartesian system, with the $Z$ axis pointing in the proton beam direction,
referred to as the ``forward direction'', and the $X$ axis pointing left
towards the centre of HERA. The coordinate origin is at the nominal
interaction point. The pseudorapidity is defined as
$\eta=-\ln\left(\tan\frac{\theta}{2}\right)$, where the polar angle, $\theta$,
is measured with respect to the proton beam direction.}, $\theta$, as viewed
from the nominal interaction point: forward 
(FCAL,~$2.6^\circ < \theta < 36.7^\circ$), barrel (BCAL,
~$36.7^\circ < \theta < 129.1^\circ$), and rear (RCAL,
~$129.1^\circ < \theta < 176.2^\circ$). Each section is sub-divided into towers
which subtend solid angles between 0.006 and 0.04 steradian. Each tower is
longitudinally segmented into an electromagnetic and one (RCAL) or two (FCAL,
BCAL) hadronic sections. The electromagnetic section of each tower is further
sub-divided transversely into two (RCAL) or four (BCAL, FCAL) cells. Under
test-beam conditions, the calorimeter single-particle relative resolutions were
$18\%/\sqrt{E\,({\rm GeV})}$ for electrons and $35\%/\sqrt{E\,({\rm GeV})}$ for
hadrons.

Tracking information is provided by the central tracking detector (CTD)
\cite{zeusctd} operating in a 1.43~T solenoidal magnetic field. The interaction
vertex is measured with a typical resolution along (transverse to) the beam
direction of 0.4~(0.1)~cm. The CTD is used to reconstruct the momenta of
tracks in the polar angle region $15^\circ < \theta < 164^\circ$. The
transverse momentum, $p_t$, resolution for full-length tracks can be
parameterised as $\sigma(p_t)/p_t=0.0058\ p_t \oplus 0.0065\oplus 0.0014/p_t$,
with $p_t$ in GeV.

The luminosity is measured using the Bethe-Heitler reaction $e^+ p
\rightarrow e^+\gamma p$~\cite{zeuslumi}. The resulting small-angle energetic
photons are measured by the luminosity monitor, a lead-scintillator
calorimeter placed in the HERA tunnel at $Z=-107$~m.
\vspace{-0.3cm}
\section{Monte Carlo models }
NC DIS events including radiative effects were simulated using the 
HERACLES~4.5.2~\cite{heracles} MC program with the DJANGO6~2.4~\cite{django}
interface to the hadronisation programs. HERACLES includes corrections for
initial- and final-state radiation, vertex and propagator terms, and
two-boson exchange. The QCD cascade is simulated using the colour-dipole model
\cite{cdm} including the LO QCD diagrams as 
implemented\footnote{A modified treatment of parton radiation at high $Q^2$
\cite{modlon} was included.}  in
ARIADNE~4.08 \cite{ariadne} and, as a systematic check of the
final results, with the MEPS model of LEPTO 6.5~\cite{lepto}. Both MC programs
use the Lund string model \cite{lund} of JETSET 7.4~\cite{pythia} for the
hadronisation. To estimate the uncertainty caused by the modelling of
hadronisation, events were also generated with the HERWIG 5.9 ~\cite{herwig}
program, in which the fragmentation is simulated according to a cluster
model \cite{cluster}.  

The ZEUS detector response was simulated with a program based on
GEANT~3.13~\cite{geant}. The generated events were passed through the simulated
detector, subjected to the same trigger requirements as the data, and
processed by the same reconstruction and offline programs.

The simulations based on ARIADNE and LEPTO give a very good description of
the shape and magnitude of the measured distributions of the inclusive event
sample as a function of $x_{Bj}$ and $Q^2$. The shape of most of the measured
distributions of the dijet sample is also adequately described by these two
programs. LEPTO gives a better description of the jet pseudorapidity
distributions, while ARIADNE gives a better description of the shape of the
dijet cross section as a function of $Q^2$. ARIADNE was used to calculate the
acceptance corrections; both programs give very similar results.

In addition, samples of events were generated without either $Z^0$ exchange or
electroweak radiative events, allowing the measured cross sections to be
corrected for these effects, which at present are not included in the NLO
QCD programs described below.
\section{NLO QCD calculations}
The NLO QCD calculations used in the analysis are based on the program DISENT
\cite{disent}. The calculations make use of a generalised version \cite{disent}
of the subtraction method \cite{terrano} and are performed in the massless
$\overline{\rm MS}$ renormalisation and factorisation schemes. The number of
flavours was set to $5$, the scales $\mu_R$ and $\mu_F$ were both set to $Q$,
and $\alpha_s(\mu_R)$ was evaluated using the two-loop formula \cite{alpworld}.
Various sets of proton PDFs, hereafter generically referred to as MBFIT
PDFs\footnote{The MBFIT1M set of proton PDFs assumes $\alpha_s(M_{Z})=0.118$.},
resulting from a recent analysis \cite{botje}, were used. This analysis provides
the covariance matrix of the fitted PDF parameters and the derivatives as a
function of $x$ and $Q^2$ of the PDFs with respect to these parameters. These
are necessary to propagate the statistical and correlated systematic
uncertainties of each data set used in the NLO DGLAP fit to the NLO QCD
differential cross sections. The DISENT predictions have been cross-checked
with results obtained with the program DISASTER++ \cite{disaster}. Agreement at
the $1-2\%$ level has been found between these programs for all the
observables discussed in this paper.

The NLO pQCD predictions to be compared with the data were corrected by a
bin-by-bin procedure for hadronisation effects according to
$d\sigma_{2+1} = d\sigma_{2+1}^{NLO} \cdot C_{\rm had}^{-1}$.
The hadronisation correction factor, $C_{\rm had}$ 
($\equiv (1+\delta_{{\rm had}})^{-1}$), was defined as the ratio of the dijet
cross sections before and after the hadronisation process, $C_{\rm had}=
d\sigma_{2+1}^{MC,{\rm partons}}/d\sigma_{2+1}^{MC,{\rm hadrons}}$. The value
of $C_{\rm had}$ was taken as the mean of the ratios obtained using the
ARIADNE and HERWIG programs, for which the predictions agree typically within
$5\%$. The hadronisation correction was below $10\%$ for most of the phase
space. These programs only include the LO matrix elements for the partonic
cross sections and higher-order contributions are simulated using various
approximations to the parton cascade. This procedure for applying hadronisation
corrections to the NLO QCD predictions was verified by checking that the shapes
of the differential dijet cross sections at the parton level are
reproduced by the NLO QCD calculations at the $10\%$ level.
\section{Event selection and jet search}
\subsection{The NC DIS data sample}
High-$Q^2$ NC DIS events were selected according to criteria similar to those
described in a recent ZEUS publication \cite{zenc99}. The events are
characterised by a high-energy isolated positron in the detector, making them
easy to distinguish from quasi-real photoproduction ($Q^2 \sim 0$) and beam-gas
interaction backgrounds.

In addition to the energy and polar angle of the positron\footnote{The positron
polar angle, $\theta_e$, was determined from the associated track if the
positron cluster was within the CTD acceptance, and otherwise from the
position of the cluster and the reconstructed vertex.}, the variables 
$\delta$, net transverse momentum, $P_T$, and total transverse energy,
$E_T$, were used for event selection, where $\delta = E-P_Z$, $E$ is the
total energy as measured in the CAL and ${\bf P}=\sum_{i} E_i {\bf r_i}$.
The sum runs over all calorimeter energy deposits $E_i$, and ${\bf r_i}$ is
a unit vector along the line joining the reconstructed vertex and the
geometric centre of cell $i$. The following criteria were applied offline: 
\begin{itemize}
\item to ensure that event quantities can be accurately determined, a
  reconstructed vertex with $-50 < Z < 50$ cm was required;
\item to suppress photoproduction events, in which the scattered positron
  escapes through the beam hole in the RCAL, $\delta$ was required to
  be greater than 38~\gev. For perfect detector resolution, $\delta$ is
  twice the positron beam energy ($55$~GeV) for fully contained DIS
  events, while, for photoproduction events, $\delta$ peaks at much lower
  values. This cut also rejected events with hard initial-state QED radiation.
  The additional requirement $\delta<65$~\gev\ removed cosmic ray background;
\item positrons were identified based on calorimeter cluster quantities
  and tracking: 
\begin{itemize}
  \item to ensure high purity, the positron was required to have an
       energy of at least 10~\gev; 
  \item to reduce background, isolated positrons were selected by requiring no
       more than 5~\gev\ in calorimeter cells not associated
       with the scattered positron in an $\eta-\phi$ cone of radius 0.8
       centred on the positron direction;
  \item each positron with $\theta_e > 17.2^\circ$ was required to
       match to a charged track of at least 5~\gev\ momentum. For positrons
       beyond the tracking acceptance ($\theta_e < 17.2^\circ$), the tracking
       requirement in the positron selection was replaced by a cut on the
       transverse momentum of the positron $p_T^e > 30$~\gev\ and by the
       requirement $\delta > 44$~\gev;
  \item a fiducial volume cut was applied to the positron position which
       excludes the upper part of the central RCAL area ($20\times 80$~cm$^2$)
       occluded by the cryogenic supply for the solenoid magnet. The transition
       regions between FCAL and BCAL, and BCAL and RCAL, corresponding to 
       polar angles of $35.6^\circ < \theta < 37.3^\circ$ and
       $128.2^\circ < \theta < 140.2^\circ$, respectively, were also excluded;
\end{itemize}
\item to reduce further the background from photoproduction, the variable
  $y=Q^2/(x_{Bj}s)$, estimated from the positron energy and angle, $y_e$,
  was required to satisfy $y_e < 0.95$;
\item the net transverse momentum, \Pt\ , is expected to be close to zero and
  was measured with an error approximately proportional to
  $\sqrt{\Et({\rm GeV})}$. To remove cosmic rays and beam-related background,
  \Pt(GeV) was required to be less than $4\sqrt{\Et\,(\gev)}$.
\end{itemize}

The kinematic variables \qq, $x_{Bj}$ and $y$ were determined using
the double angle (DA) reconstruction method~\cite{dameth} and
are referred to as $\qqda, \xda$ and
$\yda$, respectively. 
The DA method is insensitive to errors in the absolute energy scale of
the calorimeter and MC studies \cite{zenc99} have shown it to be superior to
other methods in the high-\qq\ region considered here.

\subsection{Jet search and reconstruction}
The longitudinally invariant $k_T$-cluster algorithm \cite{kt} was used
in the inclusive mode to reconstruct jets in the hadronic final state both in
data and in MC simulated events. At the detector level, the algorithm was
applied to the energy deposits in the CAL cells after excluding those
associated with the scattered positron candidate. The jet search was also
applied to the generated hadrons and to the partons in the MC samples, as well
as to the partons of the NLO QCD programs. The jet search was performed
in the Breit frame \cite{breit}, in which the exchanged virtual boson is purely
space-like with three-momentum ${\bf q} = (0,0,-Q)$. In this frame, the
selection of the jets in terms of transverse energy with respect to the
direction of the virtual boson allows a natural suppression of the
contribution due to single-jet events and strongly reduces the contamination
from the proton remnant. The boost was performed using the four-momenta of the
incoming and scattered positrons. The scattered positron four-momentum was
calculated using the polar and azimuthal angles of the positron
track\footnote{For events with $\theta_e < 17.2^\circ$, the polar and azimuthal
angles were calculated from the position of the positron calorimeter cluster
and the reconstructed vertex position.} and the DA positron energy, defined as
$E^{'}_{e,{\rm DA}}= \qqda/(2E_e(1+\cos\theta_e))$. 

The jet algorithm uses quantities defined in the Breit frame and with respect
to the direction of the incoming proton: the transverse energy, $E_{T,B}^i$,
the pseudorapidity, $\eta_B^i$, and the azimuthal angle, $\phi_B^i$, of the
object $i$. The jet variables were defined according to the Snowmass
convention \cite{snow}:
\begin{equation}
 \etjet = \sum_i E_{T,B}^i \; , \;
 \etajet = \frac{\sum_i E_{T,B}^i \; \eta_B^i}{\etjet} \; , \;
 \phijet = \frac{\sum_i E_{T,B}^i \; \phi_B^i}{\etjet}. \nonumber
\end{equation}

The comparison of the reconstructed jet variables between jets of hadrons
and jets of CAL cells in MC generated events showed no significant
systematic shift in the angular variables $\etajet$ and $\phijet$.
However, the jet transverse energy as measured by the CAL under-estimated
that of the jet of hadrons by an average of $\approx 15\%$. This effect is due
mainly to energy losses in the inactive material in front of the CAL and was
corrected \cite{zeusjs,etthesis} using the samples of MC generated events.

\subsection{Selected kinematic region}
The differential cross sections and dijet fraction presented in Section~7 are
quoted for the DIS kinematic region defined by $470 < Q^2 < 20000$~GeV$^2$ and
$0 < y <  1$, and for the jet selection criteria $E_{T,B}^{{\rm jet},M}>8$~GeV,
$E_{T,B}^{{\rm jet},m} > 5$~GeV and $-1 < \eta^{{\rm jet}}_{Lab} < 2$, 
where $E_{T,B}^{{\rm jet},M}$ ($E_{T,B}^{{\rm jet},m}$) is the transverse
energy of the jet in the Breit frame with the highest (second highest)
transverse energy in the event. Only events with exactly two jets passing the
above selection cuts were used to compute the cross sections. The final
sample contains $1637$ dijet events. The restriction
to a high-$Q^2$ region was chosen to avoid the large
renormalisation-scale dependence of the NLO QCD dijet cross sections at lower
$Q^2$ ($20-50\%$ in the region $10 < Q^2 < 100$~GeV$^2$) \cite{etthesis}, to
reduce the uncertainty due to the proton PDFs (in particular the gluon
density), and to improve the reconstruction of the boost to the
Breit frame. The asymmetric cuts on the $\etjet$ of the jets avoided infrared
sensitive regions where the behaviour of the cross section as predicted by the
NLO QCD programs is unphysical \cite{klasen}.

\section{Results}
\subsection{Dijet differential cross sections}
\begin{figure}
\vspace{-1.0cm}
\centerline{\epsfig{figure=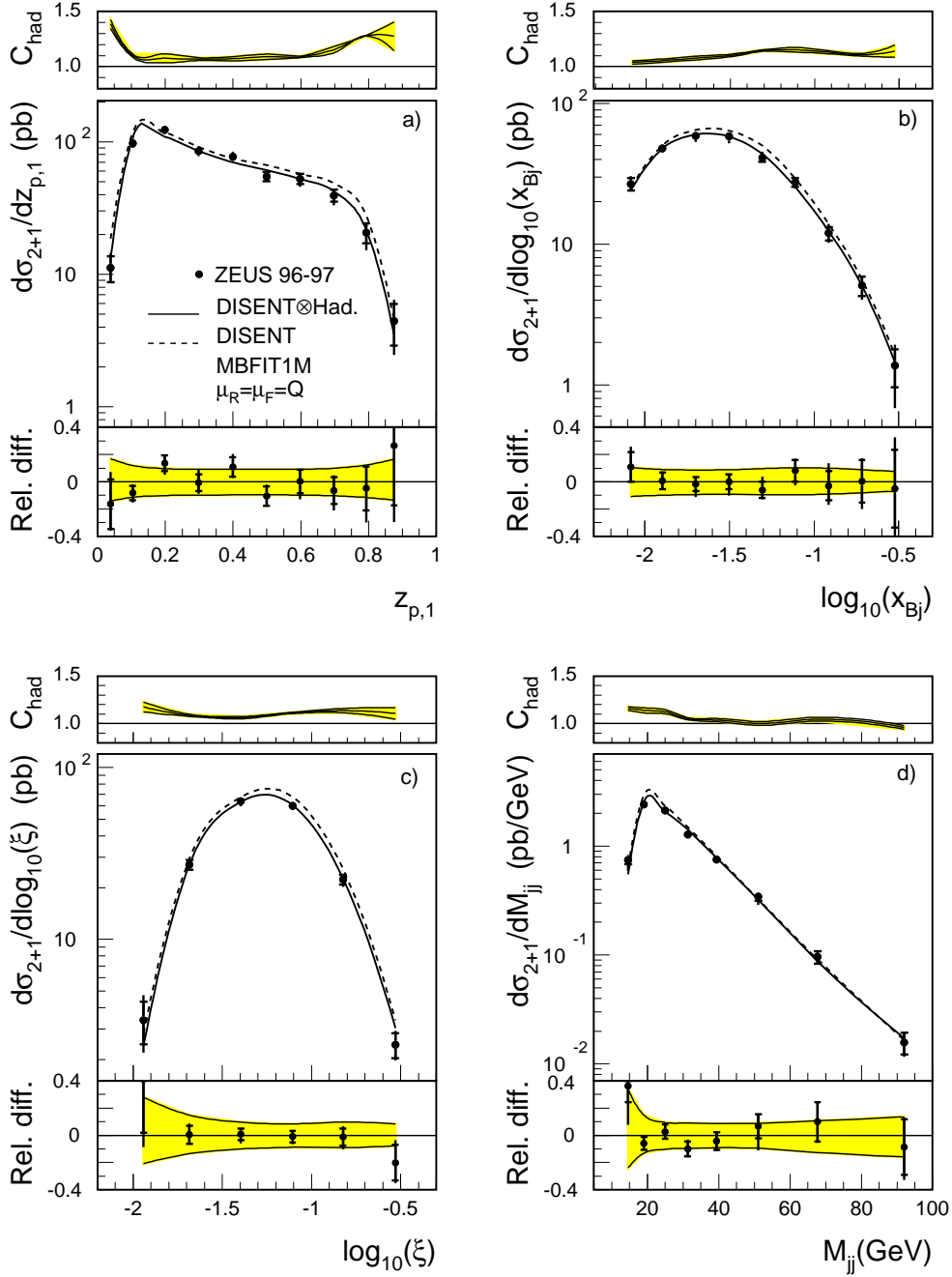,width=13cm}}
\vspace{-2.4cm}
\caption{\label{fig2}
{\small The measured differential dijet cross sections in NC DIS as functions of
a) $z_{p,1}$, b) $\log_{10}x_{Bj}$, c) $\log_{10}\xi$ and d) dijet
invariant mass $M_{jj}$. The inner error bars represent the statistical errors
of the data. The outer error bars show the statistical errors and systematic
uncertainties $-$~except those associated with the uncertainty in the
absolute energy scale of the jets~$-$ added in quadrature. For comparison,
pure NLO QCD calculations (dashed lines) and NLO QCD calculations corrected
for hadronisation effects (solid lines), obtained using the proton MBFIT
PDFs and $\mu_R=\mu_F=Q$, are shown. The relative differences of the measured
differential cross sections over the NLO QCD predictions corrected for
hadronisation effects are shown underneath each plot; the shaded band
represents the uncertainty of the QCD calculation (see text). The
hadronisation correction ($C_{\rm had}$) together with its uncertainty are
shown above each plot.}}
\end{figure}

\begin{figure}
\vspace{+0cm}
\centerline{\epsfig{figure=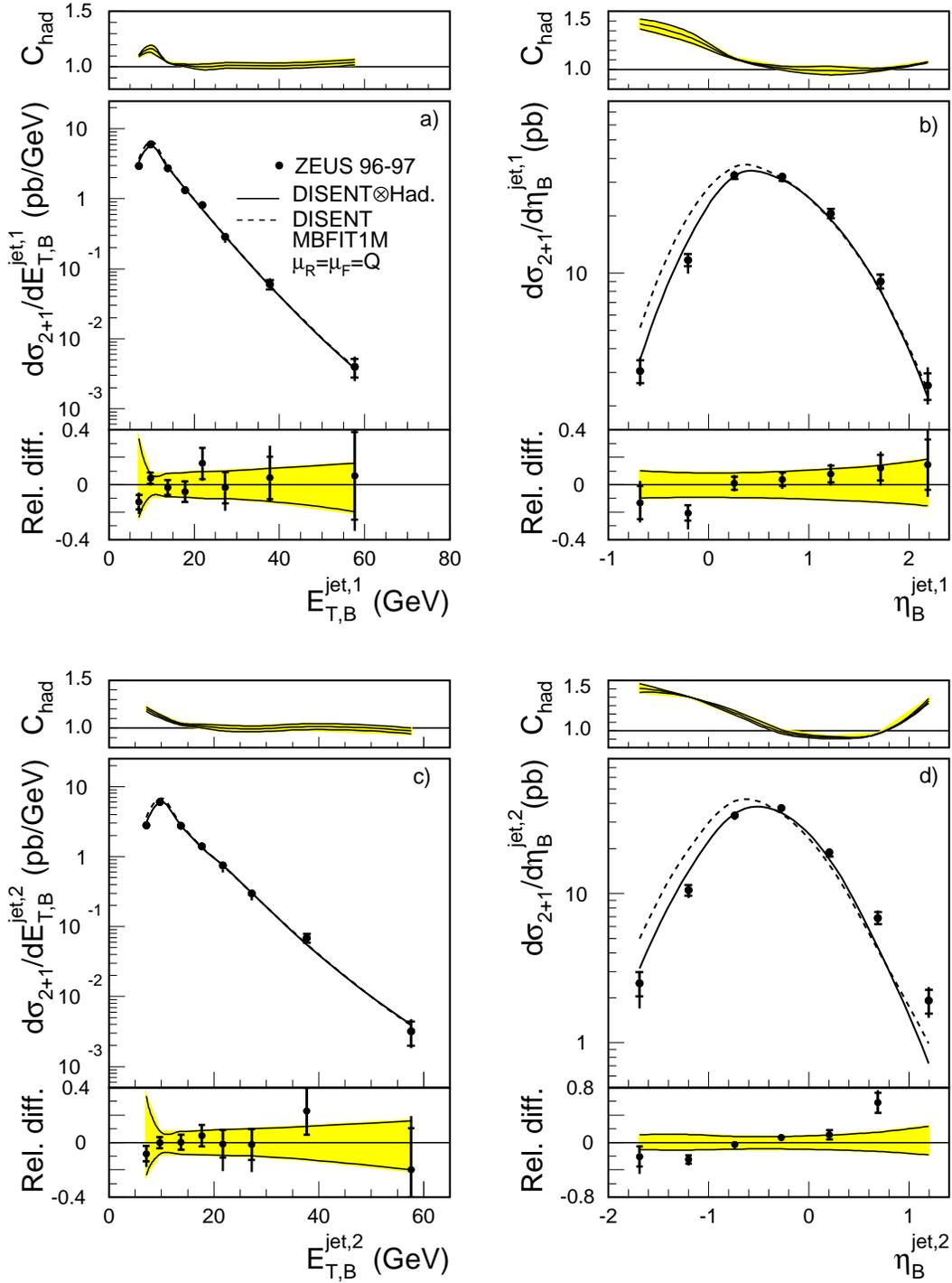,width=14cm}}
\vspace{-2.0cm}
\caption{\label{fig3}
{\small The measured differential dijet cross sections in NC DIS as functions of
a) $E_{T,B}^{{\rm jet},1}$, b) $\eta_B^{{\rm jet},1}$,
c) $E_{T,B}^{{\rm jet},2}$ and d) $\eta_B^{{\rm jet},2}$. Other details are
as described in the caption to Fig.~\ref{fig2}.}}
\end{figure}

\begin{figure}
\vspace{0cm}
\centerline{\epsfig{figure=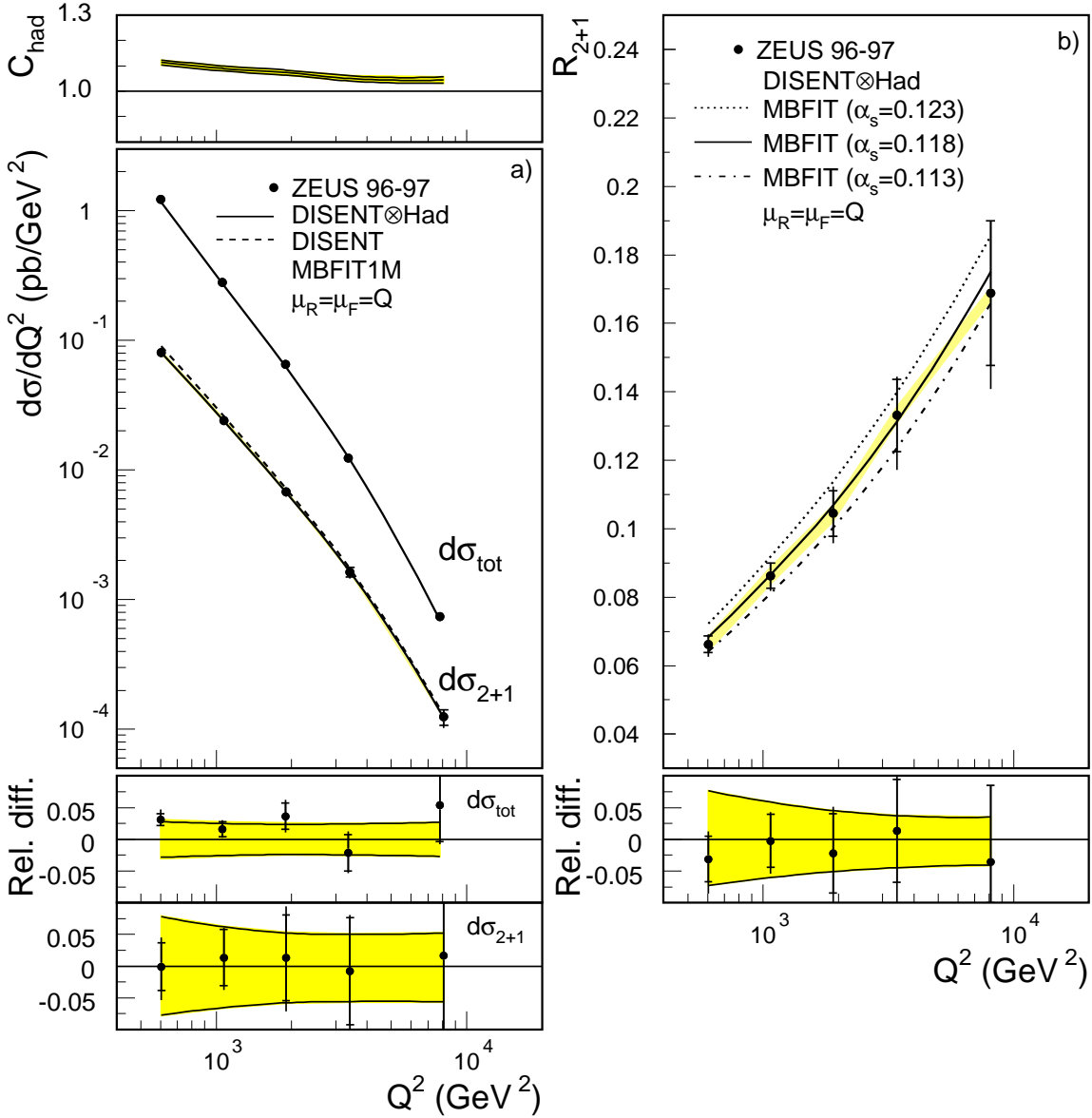,width=17cm}}
\vspace{0.0cm}
\caption{\label{fig4}
{\small a) The measured inclusive ($\sq2$) and dijet ($\sdijq2$) differential 
cross sections in NC DIS as a function of $Q^2$. The hadronisation correction
($C_{\rm had}$), shown above the figure, refers to the dijet cross section.
b) The dijet fraction, $\rat21$, in NC DIS as a function of $Q^2$. The light
shaded band displays the uncertainty due to the absolute energy scale of the
jets. For comparison, the QCD predictions using MBFIT proton PDFs determined
assuming $\alpha_s(M_{Z})=0.113$ and $0.123$ \cite{botje} are also shown. The
bands showing the theoretical uncertainty on the cross sections and dijet
fraction do not include the uncertainty associated with $\alpha_s(M_{Z})$.
Other details are as described in the caption to Fig.~\ref{fig2}.}}
\end{figure}

The cross sections measured in the kinematic region defined in the previous 
section were corrected for detector effects, QED radiative effects and
$Z^{0}$-exchange processes. In the presented dijet cross sections, the two
jets were ordered according to decreasing pseudorapidity in the Breit frame 
($\eta_B^{{\rm jet},1}>\eta_B^{{\rm jet},2}$). The measurements\footnote{Tables
of the results and their associated uncertainties are available in
electronic form from the ZEUS WWW site at
http://www-zeus.desy.de/zeus\_papers/zeus\_papers.html.
They can also be obtained by contacting the authors.}
of the differential dijet cross sections are presented in Figs.~\ref{fig2} and
\ref{fig3}. The inclusive and dijet cross sections, as well as the dijet
fraction, $\rat21 \equiv (\sdijq2)/(\sq2)$,
are presented as a function of $Q^2$ in Fig.~\ref{fig4}.

In Figs.~\ref{fig2} to \ref{fig4}, the measured differential dijet cross 
sections and dijet fraction are compared to DISENT NLO QCD predictions,
corrected for hadronisation effects. The hadronisation correction,
$C_{\rm had}$, as well as its uncertainty, and the NLO QCD cross sections
without hadronisation corrections are shown in Figs.~\ref{fig2} to \ref{fig4}. 
The QCD predictions, which assume $\alpha_s(M_{Z})=0.118$, provide a good
overall description of both the shape and the magnitude of the measured cross
sections. The only case where the agreement between data and NLO QCD is not
good is the cross section as a function of the pseudorapidity of the most
backward jet in the Breit frame (see Fig.~\ref{fig3}d), where the data lie
above the theoretical predictions in the region $\eta_B^{{\rm jet},2} > 0.5$.
However, hadronisation effects are particularly large for this variable and
vary more rapidly than for other variables. The impact of such a discrepancy
in the determination of $\alpha_s(M_{Z})$ is minor since the contribution to
the total dijet cross section from the region $\eta_B^{{\rm jet},2}>0.5$ is
small. The dijet fraction increases with increasing $Q^2$ due to phase-space
effects. For the cross sections as a function of the jet transverse energies
and $Q^2$, there is agreement at the $\approx  10\%$ level between data and
theory over four orders of magnitude, demonstrating the validity of the
description of the dynamics of dijet production by the NLO QCD hard processes.

\subsection{Experimental and theoretical uncertainties}
A detailed study of each of the main sources contributing to the systematic
uncertainties of the measurements has been performed \cite{etthesis}. These
sources, for which a typical value of the systematic uncertainty in the dijet
cross section $\sdijq2$ is indicated in parentheses, are listed below:
\begin{itemize}
 \item uncertainties in the positron identification efficiency and in the
       positron energy-scale \cite{zenc99} ($1\%$);
 \item uncertainties in the reconstruction of the boost to the Breit frame
      ($1\%$);
 \item use of the LEPTO program instead of ARIADNE to evaluate the acceptance
      corrections to the observed dijet distributions ($2\%$); 
 \item uncertainties in the simulation of the trigger and in the cuts used to
      select the data ($2\%$);
 \item uncertainty of $\pm 2\%$ in the absolute jet energy scale of the CAL
      ($\pm 3.5\%$) obtained by examining the transverse momentum balance between the
      scattered positron and the jets.
\end{itemize}
The last uncertainty is the dominant source of experimental systematic
uncertainty and is strongly correlated between measurements at different
points. It is shown as a light shaded band in Fig.~\ref{fig4}b. In addition,
there is an overall normalisation uncertainty of $1.6\%$ from the luminosity
determination, which is not included in the figures.

\begin{figure}
\vspace{-2cm}
\centerline{\epsfig{figure=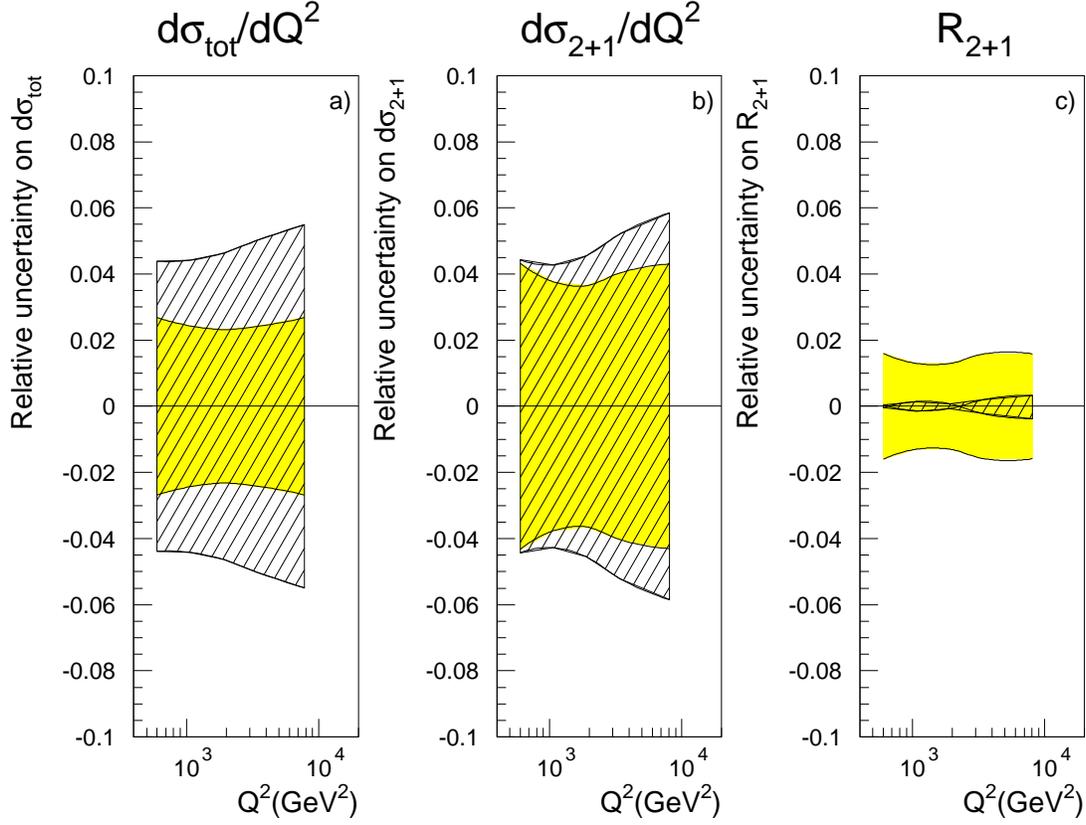,width=16cm}}
\vspace{+1cm}
\caption{\label{fig1}
{\small The relative uncertainty on a) the inclusive, b) dijet differential
cross sections and c) the dijet fraction, due to the statistical and systematic
experimental uncertainties of each data set used in the determination of the
MBFIT PDFs. The shaded (hatched) bands indicate the uncertainties obtained
(not) taking into account the correlations among the PDFs parameters.}}
\end{figure}

The NLO QCD predictions for the dijet cross sections are affected by the
following:
\begin{itemize}
 \item uncertainties in the hadronisation correction, which were estimated as
      half the spread between the $C_{\rm had}$ values obtained using the
      ARIADNE (string) and HERWIG (cluster) models ($1\%$);
 \item uncertainties due to terms beyond NLO, which were estimated by varying
      $\mu_R$ between $Q/2$ and $2Q$, keeping $\mu_F$ fixed at $Q$ ($6\%$);
 \item uncertainty in the value of $\alpha_s(M_{Z})$, which was estimated by
      repeating the calculations using MBFIT proton PDFs determined assuming
      $\alpha_s(M_{Z})=0.113$ and $0.123$ \cite{botje} ($6\%$);
 \item uncertainties in the proton PDFs. The uncertainty due to the statistical
      and correlated systematic experimental uncertainties of each data set
      used in the determination of the MBFIT PDFs was calculated making use of
      the covariance matrix provided \cite{botje}. To estimate the
      uncertainties on the NLO QCD differential cross sections due to the
      theoretical uncertainties affecting the extraction of the NLO PDFs, the
      calculation of all the differential cross sections was repeated
      using a number of different MBFIT PDFs obtained under different
      theoretical assumptions in the DGLAP fit \cite{botje}. This uncertainty
      was added in quadrature to that estimated above to give the total
      PDF-related uncertainty. The importance of taking the correlations among
      the PDF parameters into account is illustrated in Fig.~\ref{fig1}, which
      shows, as a function of $Q^2$, the relative uncertainties arising from
      the PDFs on the inclusive and dijet cross sections, and the dijet
      fraction. The uncertainties were
      obtained by including or ignoring the information provided by the
      covariance matrix mentioned above. The small uncertainty on $\rat21$ due
      to the PDFs makes this observable particularly suitable for extracting 
      $\as$ ($2.5\%$ on $\sq2$, $4\%$ on $\sdijq2$ and $1.5\%$ on $\rat21$).
\end{itemize}
The total theoretical uncertainty was obtained by adding in quadrature
the individual uncertainties listed above.

\subsection{Determination of $\alpha_s$}
The measured dijet fraction as a function of $Q^2$, $\rat21$, was used to
determine $\alpha_s(M_{Z})$. The sensitivity of the measurements to the value
of $\alpha_s(M_{Z})$ is illustrated in Fig.~\ref{fig4}b, which compares the
measured $\rat21$ with the NLO QCD calculations for three values of
$\alpha_s(M_{Z})$.

The procedure to determine $\alpha_s(M_{Z})$ was as follows:
\begin{itemize}
 \item NLO QCD calculations of $\rat21$ were performed for three sets of
      MBFIT proton PDFs obtained assuming $\alpha_s(M_{Z})=0.113$, 0.118 and
      0.123, respectively \cite{botje}. The value of $\alpha_s(M_{Z})$ used in
      each partonic cross-section calculation was that associated with the
      corresponding set of PDFs;
 \item for each bin, $i$, in $Q^2$, the NLO QCD calculations mentioned above,
      corrected for hadronisation effects, were used to parameterise the
      $\alpha_s$ dependence of the dijet fraction according to the functional 
      form:
\begin{equation}
 R^i_{2+1}(\alpha_s(M_{Z}))=A^i_1\cdot \alpha_s(M_{Z})+A^i_2\cdot \alpha^2_s(M_{Z}).
 \label{param}
\end{equation}
      This parameterisation allows a simple description of the
      $\alpha_s$ dependence of $\rat21$ over the entire $\alpha_s$ range
      spanned by the MBFIT PDF sets, while using only three NLO calculations
      of the dijet fraction;
 \item the value of $\alpha_s(M_{Z})$ was then determined by a $\chi^2$-fit
      of Eq.~(\ref{param}) to the measured $\rat21$ values.
\end{itemize}

This procedure correctly handles the complete $\alpha_s$-dependence of
the NLO differential cross sections (the explicit dependence coming from the
partonic cross sections and the implicit one coming from the PDFs) in the
fit, while preserving the correlation between $\alpha_s$ and the PDFs. Its
stability was checked with respect to variations in the PDFs and $\alpha_s$,
as well as to alternative parameterisations of $R_{2+1}(Q^2,\alpha_s(M_{Z}))$. 

Taking into account only the statistical errors on the measured dijet fraction,
$\alpha_s(M_{Z})$ is found to be $0.1166 \pm 0.0019 \; {\rm (stat.)}$.
The uncertainty on the value of $\alpha_s(M_{Z})$ due to the experimental 
systematic uncertainties of the measured dijet fraction was evaluated by
repeating the analysis above for each systematic check. For the ratio
$\rat21$, some of the systematic uncertainties largely cancel. The total
experimental systematic uncertainty on the value of $\alpha_s(M_{Z})$ is 
$^{+ 0.0024}_{- 0.0033}$. The largest uncertainty is that on the jet energy
scale.

The following sources of theoretical uncertainties (evaluated as described
in Section~7.2) and cross-checks on the extracted value of $\alpha_s(M_{Z})$
were considered \cite{etthesis}:
\begin{itemize}
 \item terms beyond NLO: $\Delta \alpha_s(M_{Z}) = {}^{+ 0.0055}_{- 0.0042}$;
 \item uncertainties in the proton PDFs: 
       $\Delta \alpha_s(M_{Z}) = {}^{+ 0.0012}_{- 0.0011}$;
 \item hadronisation effects: $\Delta \alpha_s(M_{Z}) = \pm 0.0005$;
 \item the dependence on the use of a renormalisation scale that involves the
      jet variables. The analysis was repeated with $\mu_R=
      (E_{T,B}^{{\rm jet},1}+E_{T,B}^{{\rm jet},2})$. The result is
      $\alpha_s(M_{Z}) = 0.1125 \pm 0.0018 \; {\rm (stat.)}$, which is almost
      equal to that obtained with $\mu_R=Q/2$. This source of uncertainty
      was therefore neglected;
 \item the fit procedure. This was cross-checked by repeating the 
      $\alpha_s(M_{Z})$ determination using the five sets of proton PDFs of
      the CTEQ4 ``A-series'' \cite{cteq4} and the three MRST sets, central,
      $\alpha_s \uparrow\uparrow$ and $\alpha_s \downarrow\downarrow$
      \cite{mrst}. The results are in good agreement with the central value
      determined above. This source of uncertainty
      was therefore neglected
\end{itemize}
The total theoretical uncertainty of $^{+ 0.0057}_{- 0.0044}$ was obtained by
adding in quadrature the uncorrelated uncertainties on $\alpha_s(M_{Z})$ due
to the first three items mentioned above.

The value of $\alpha_s(M_{Z})$ as determined from the measured $\rat21$ is
therefore:
\begin{equation}
 \alpha_s(M_{Z}) = 0.1166 \pm 0.0019 \; {\rm (stat.)}
                          ^{+ 0.0024}_{- 0.0033} \; {\rm (exp.)}
                          ^{+ 0.0057}_{- 0.0044} \; {\rm (th.)}. \nonumber
\end{equation}
This result is consistent with the current PDG world average, 
$\alpha_s(M_{Z})=0.1181 \pm 0.0020$ \cite{alpworld}, a review from Bethke \cite{bethke}, and recent
determinations by the H1 Collaboration \cite{h1alp}.

\subsection{The energy scale dependence of $\alpha_s$}
\begin{figure}
\vspace{-2cm}
\centerline{\epsfig{figure=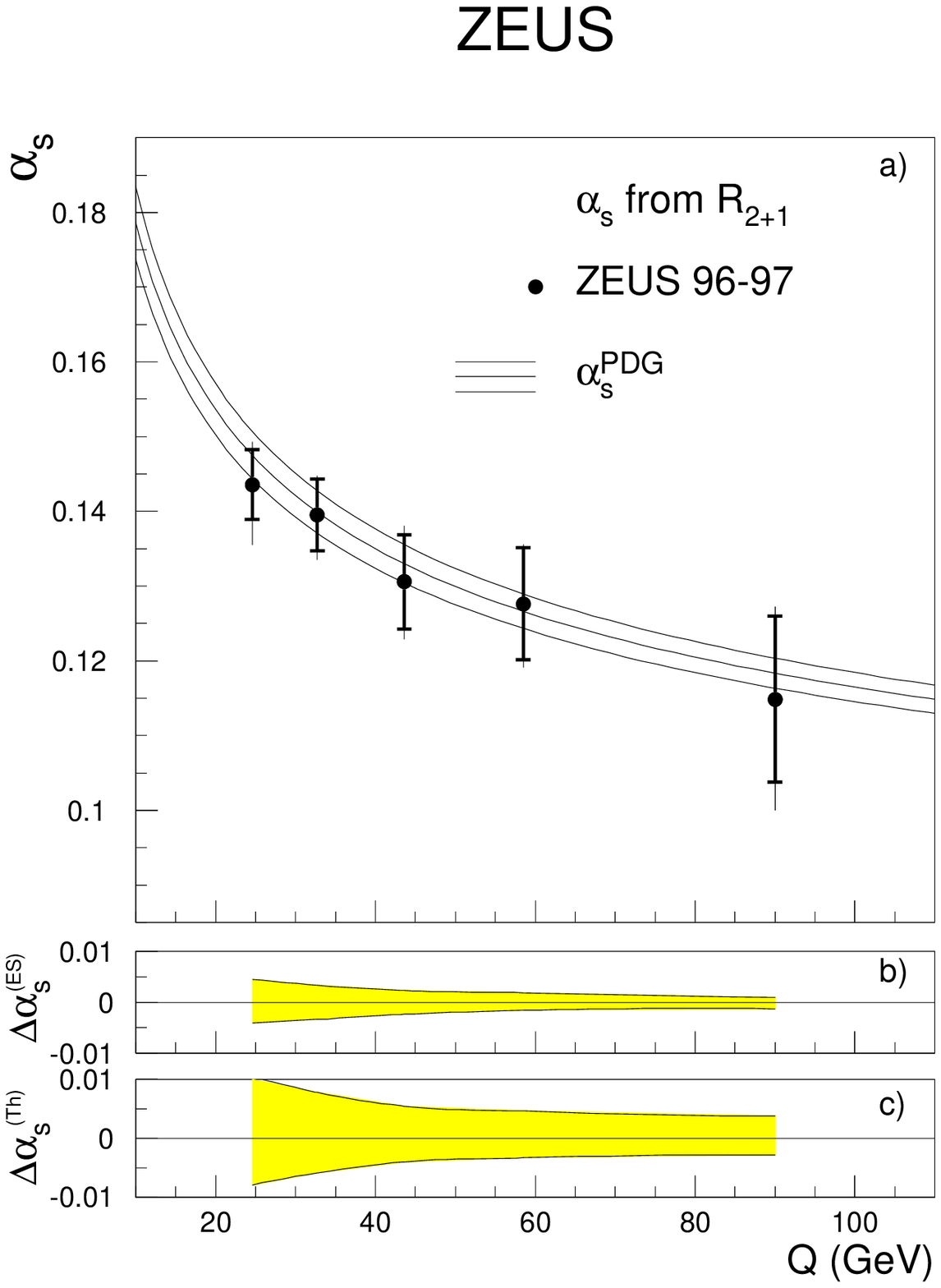,width=15cm}}
\vspace{-1cm}
\caption{\label{fig5}
{\small a) The $\alpha_s(Q)$ values determined from the QCD fit of the measured
dijet fraction, $\rat21$, as a function of $Q$. The inner error bars represent
the statistical errors of the data. The outer error bars show the statistical
errors and systematic uncertainties $-$~except those associated with the
uncertainty in the absolute energy scale of the jets~$-$ added in quadrature. 
The three curves indicate the renormalisation group predictions obtained from
the PDG $\alpha_s(M_Z)$ value and its associated uncertainty. b) The
uncertainty on $\alpha_s$ due to the absolute energy scale of the jets,
$\Delta\alpha_s^{ES}$. c) The total theoretical uncertainty associated with the
determination of $\alpha_s$, $\Delta \alpha_s^{\rm Th}$.}}
\end{figure}

A consistency test for the scale dependence of the renormalised strong coupling constant
predicted by the renormalisation group equation was carried out by repeating the QCD fit of the dijet
fraction in five $Q^2$ bins. The principle of the fit is the
same as outlined above, with the only difference being that the $\alpha_s$
dependence of the dijet fraction in Eq.~(\ref{param}) was parameterised not in
terms of $\alpha_s(M_{Z})$ but in terms of $\alpha_s(\langle Q \rangle)$,
where $\langle Q \rangle$ is the mean value of $Q$ in each bin. The measured 
$\alpha_s(\langle Q \rangle)$ values, with their experimental and theoretical
systematic uncertainties estimated as for $\alpha_s(M_{Z})$, are shown in
Fig.~\ref{fig5}. The measurements are compared with the renormalisation group
predictions obtained from the PDG $\alpha_s(M_Z)$ value and its associated
uncertainty. The values are in good agreement with the predicted running of the
strong coupling constant over a large range in $Q$.

\section{Summary}
Differential dijet cross sections have been measured in neutral current
deep inelastic $e^+p$ scattering for $470 < Q^2 < 20000$~GeV$^2$ with the
ZEUS detector at HERA. These measurements, which use events with only two
jets, were performed in a kinematic region where both theoretical and
experimental uncertainties are small.
Next-to-leading-order QCD calculations give a good description of the
shape and magnitude of the kinematic distributions and cross sections.
For the cross sections as a function of the jet transverse energies and $Q^2$,
there is agreement at the $\approx  10\%$ level between data and theory
over four orders of magnitude, demonstrating the validity of the description
of the dynamics of dijet production by the NLO QCD hard processes. A QCD fit
of the measured dijet fraction as a function of $Q^2$ provides both a precise
determination of the strong coupling constant and a test of its energy-scale
dependence. A comprehensive analysis of the uncertainties of the calculations
has been carried out, which takes into account the dependence of the proton
PDFs on the assumed value of $\alpha_s$ and the statistical and correlated
systematic uncertainties from each data set used in the determination of the
proton PDFs. The value of $\alpha_s(M_{Z})$ as determined by fitting the 
next-to-leading-order QCD calculations to the measured dijet fraction is
\begin{equation}
       \alpha_s(M_{Z}) = 0.1166 \pm 0.0019 \; {\rm (stat.)}
                         ^{+ 0.0024}_{- 0.0033} \; {\rm (exp.)}
                         ^{+ 0.0057}_{- 0.0044} \; {\rm (th.)}\; , \nonumber
\end{equation}
in good agreement with the world average \cite{alpworld}. The value of 
$\alpha_s$ as a function of $Q$ is in good agreement, over a wide range of
$Q$, with the running of $\alpha_s$ as predicted by QCD.

\vspace{0.5cm}
\noindent {\Large\bf Acknowledgements}
\vspace{0.3cm}

We thank the DESY Directorate for their strong support and encouragement.
The remarkable achievements of the HERA machine group were essential for
the successful completion of this work and are greatly appreciated. We
would like to thank D. Graudenz, M. Seymour and H. Spiesberger for valuable
discussions and help in running their NLO programs.


\setcounter{secnumdepth}{0} 


%
%

\newpage
\clearpage
 \begin{table}
 \begin{center}
 \begin{tabular}{|c|c|c|c|c|}
 \hline
 \hline
 $z_{p,1}$ range &                                                         
 $d\sigma_{2+1}/dz_{p,1}$~[pb] &                                           
 $\Delta_{\rm stat}$ &
 $\Delta_{\rm syst}$ &
 $\Delta_{\rm ES}$ \\
 \hline
 \hline
 0.00 --  0.05 &   1.118$\; \cdot \; 10^{ 1}$ & $(\pm   0.244)\cdot 10^{ 1}$ & $(^{+   0.203}_{-  0.097})\cdot 10^{ 1}$ & $(^{+   0.089}_{-  0.089})\cdot 10^{ 1}$ \\
 \hline
 0.05 --  0.15 &   9.777$\; \cdot \; 10^{ 1}$ & $(\pm   0.555)\cdot 10^{ 1}$ & $(^{+   0.365}_{-  0.519})\cdot 10^{ 1}$ & $(^{+   0.365}_{-  0.400})\cdot 10^{ 1}$ \\
 \hline
 0.15 --  0.25 &   1.233$\; \cdot \; 10^{ 2}$ & $(\pm   0.063)\cdot 10^{ 2}$ & $(^{+   0.024}_{-  0.068})\cdot 10^{ 2}$ & $(^{+   0.038}_{-  0.033})\cdot 10^{ 2}$ \\
 \hline
 0.25 --  0.35 &   8.469$\; \cdot \; 10^{ 1}$ & $(\pm   0.520)\cdot 10^{ 1}$ & $(^{+   0.625}_{-  0.482})\cdot 10^{ 1}$ & $(^{+   0.207}_{-  0.158})\cdot 10^{ 1}$ \\
 \hline
 0.35 --  0.45 &   7.738$\; \cdot \; 10^{ 1}$ & $(\pm   0.505)\cdot 10^{ 1}$ & $(^{+   0.425}_{-  0.268})\cdot 10^{ 1}$ & $(^{+   0.170}_{-  0.144})\cdot 10^{ 1}$ \\
 \hline
 0.45 --  0.55 &   5.473$\; \cdot \; 10^{ 1}$ & $(\pm   0.438)\cdot 10^{ 1}$ & $(^{+   0.288}_{-  0.145})\cdot 10^{ 1}$ & $(^{+   0.125}_{-  0.133})\cdot 10^{ 1}$ \\
 \hline
 0.55 --  0.65 &   5.259$\; \cdot \; 10^{ 1}$ & $(\pm   0.453)\cdot 10^{ 1}$ & $(^{+   0.156}_{-  0.525})\cdot 10^{ 1}$ & $(^{+   0.138}_{-  0.148})\cdot 10^{ 1}$ \\
 \hline
 0.65 --  0.75 &   3.938$\; \cdot \; 10^{ 1}$ & $(\pm   0.415)\cdot 10^{ 1}$ & $(^{+   0.240}_{-  0.451})\cdot 10^{ 1}$ & $(^{+   0.167}_{-  0.135})\cdot 10^{ 1}$ \\
 \hline
 0.75 --  0.85 &   2.064$\; \cdot \; 10^{ 1}$ & $(\pm   0.349)\cdot 10^{ 1}$ & $(^{+   0.076}_{-  0.416})\cdot 10^{ 1}$ & $(^{+   0.135}_{-  0.137})\cdot 10^{ 1}$ \\
 \hline
 0.85 --  0.95 & \hspace*{-1.00cm}   4.436 & \hspace*{-1.00cm} $\pm   1.537$ & \hspace*{-1.00cm} $^{+   1.261}_{-  1.232}$ & \hspace*{-1.00cm} $^{+   0.522}_{-  0.325}$ \\
 \hline
 \end{tabular}
 \end{center}
 \caption{\label{table1} The differential dijet cross section $d\sigma_{2+1}/dz_{p,1}$.
 For each bin in $z_{p,1}$, the measured cross section, the statistical uncertainty, 
 $\Delta_{\rm stat}$, and the systematic uncertainty (not) associated with the energy
 scale of the jets, $\Delta_{\rm ES}$ ($\Delta_{\rm syst}$), are given. 
 The overall normalisation uncertainty of $1.6\%$ due to the luminosity
 determination is not included.}
 \end{table}
 
 \begin{table}
 \begin{center}
 \begin{tabular}{|c|c|c|c|c|}
 \hline
 \hline
 $\log_{10}(x_{Bj})$ range &                                           
 $d\sigma_{2+1}/d\log_{10}(x_{Bj})$~[pb] &                             
 $\Delta_{\rm stat}$ &
 $\Delta_{\rm syst}$ &
 $\Delta_{\rm ES}$ \\
 \hline
 \hline
-2.20 -- -2.00 &   2.681$\; \cdot \; 10^{ 1}$ & $(\pm   0.267)\cdot 10^{ 1}$ & $(^{+   0.241}_{-  0.093})\cdot 10^{ 1}$ & $(^{+   0.072}_{-  0.064})\cdot 10^{ 1}$ \\
 \hline
-2.00 -- -1.80 &   4.794$\; \cdot \; 10^{ 1}$ & $(\pm   0.286)\cdot 10^{ 1}$ & $(^{+   0.139}_{-  0.118})\cdot 10^{ 1}$ & $(^{+   0.114}_{-  0.115})\cdot 10^{ 1}$ \\
 \hline
-1.80 -- -1.60 &   5.902$\; \cdot \; 10^{ 1}$ & $(\pm   0.312)\cdot 10^{ 1}$ & $(^{+   0.143}_{-  0.472})\cdot 10^{ 1}$ & $(^{+   0.162}_{-  0.166})\cdot 10^{ 1}$ \\
 \hline
-1.60 -- -1.40 &   5.821$\; \cdot \; 10^{ 1}$ & $(\pm   0.319)\cdot 10^{ 1}$ & $(^{+   0.110}_{-  0.497})\cdot 10^{ 1}$ & $(^{+   0.175}_{-  0.180})\cdot 10^{ 1}$ \\
 \hline
-1.40 -- -1.20 &   4.105$\; \cdot \; 10^{ 1}$ & $(\pm   0.263)\cdot 10^{ 1}$ & $(^{+   0.341}_{-  0.141})\cdot 10^{ 1}$ & $(^{+   0.146}_{-  0.138})\cdot 10^{ 1}$ \\
 \hline
-1.20 -- -1.00 &   2.750$\; \cdot \; 10^{ 1}$ & $(\pm   0.203)\cdot 10^{ 1}$ & $(^{+   0.112}_{-  0.181})\cdot 10^{ 1}$ & $(^{+   0.095}_{-  0.070})\cdot 10^{ 1}$ \\
 \hline
-1.00 -- -0.80 &   1.195$\; \cdot \; 10^{ 1}$ & $(\pm   0.131)\cdot 10^{ 1}$ & $(^{+   0.155}_{-  0.112})\cdot 10^{ 1}$ & $(^{+   0.041}_{-  0.034})\cdot 10^{ 1}$ \\
 \hline
-0.80 -- -0.60 & \hspace*{-1.00cm}   5.078 & \hspace*{-1.00cm} $\pm   0.793$ & \hspace*{-1.00cm} $^{+   0.323}_{-  0.670}$ & \hspace*{-1.00cm} $^{+   0.176}_{-  0.105}$ \\
 \hline
-0.60 -- -0.40 & \hspace*{-1.00cm}   1.373 & \hspace*{-1.00cm} $\pm   0.414$ & \hspace*{-1.00cm} $^{+   0.363}_{-  0.552}$ & \hspace*{-1.00cm} $^{+   0.042}_{-  0.023}$ \\
 \hline
 \end{tabular}
 \end{center}
 \caption{\label{table2} The differential dijet cross section $d\sigma_{2+1}/d\log_{10}(x_{Bj})$.
 Other details are as described in the caption to Table~\ref{table1}.}
 \end{table}

\newpage
\clearpage
 \begin{table}
 \begin{center}
 \hspace*{0.3cm}
 \begin{tabular}{|c|c|c|c|c|}
 \hline
 \hline
 $\log_{10}(\xi)$ range &                                              
 $d\sigma_{2+1}/d\log_{10}(\xi)$~[pb] &                                
 $\Delta_{\rm stat}$ &
 $\Delta_{\rm syst}$ &
 $\Delta_{\rm ES}$ \\
 \hline
 \hline
  -2.1875 --   -1.8750 & \hspace*{-1.00cm}   3.386 & \hspace*{-1.00cm} $\pm   0.939$ & \hspace*{-1.00cm} $^{+   0.970}_{-  0.736}$ & \hspace*{-1.00cm} $^{+   0.093}_{-  0.078}$ \\
 \hline
  -1.8750 --   -1.5625 &   2.726$\; \cdot \; 10^{ 1}$ & $(\pm   0.179)\cdot 10^{ 1}$ & $(^{+   0.126}_{-  0.045})\cdot 10^{ 1}$ & $(^{+   0.021}_{-  0.037})\cdot 10^{ 1}$ \\
 \hline
  -1.5625 --   -1.2500 &   6.379$\; \cdot \; 10^{ 1}$ & $(\pm   0.266)\cdot 10^{ 1}$ & $(^{+   0.085}_{-  0.334})\cdot 10^{ 1}$ & $(^{+   0.147}_{-  0.136})\cdot 10^{ 1}$ \\
 \hline
  -1.2500 --   -0.9375 &   5.997$\; \cdot \; 10^{ 1}$ & $(\pm   0.260)\cdot 10^{ 1}$ & $(^{+   0.156}_{-  0.224})\cdot 10^{ 1}$ & $(^{+   0.199}_{-  0.189})\cdot 10^{ 1}$ \\
 \hline
  -0.9375 --   -0.6250 &   2.232$\; \cdot \; 10^{ 1}$ & $(\pm   0.142)\cdot 10^{ 1}$ & $(^{+   0.105}_{-  0.154})\cdot 10^{ 1}$ & $(^{+   0.126}_{-  0.108})\cdot 10^{ 1}$ \\
 \hline
  -0.6250 --   -0.3125 & \hspace*{-1.00cm}   2.434 & \hspace*{-1.00cm} $\pm   0.400$ & \hspace*{-1.00cm} $^{+   0.305}_{-  0.229}$ & \hspace*{-1.00cm} $^{+   0.177}_{-  0.123}$ \\
 \hline
 \end{tabular}
 \end{center}
 \caption{\label{table3} The differential dijet cross section $d\sigma_{2+1}/d\log_{10}(\xi)$.
 Other details are as described in the caption to Table~\ref{table1}.}
 \vspace{0.3cm}
 \end{table}

 \begin{table}
 \begin{center}
 \hspace*{-0.1cm}
 \begin{tabular}{|c|c|c|c|c|}
 \hline
 \hline
 $M_{jj}$ range [GeV]&                                                      
 $d\sigma_{2+1}/dM_{jj}$~[pb/GeV] &                                    
 $\Delta_{\rm stat}$ &
 $\Delta_{\rm syst}$ &
 $\Delta_{\rm ES}$ \\
 \hline
 \hline
  10.00 --   16.00 &   7.462$\; \cdot \; 10^{-1}$ & $(\pm   0.642)\cdot 10^{-1}$ & $(^{+   0.562}_{-  1.434})\cdot 10^{-1}$ & $(^{+   0.016}_{-  0.005})\cdot 10^{-1}$ \\
 \hline
  16.00 --   22.00 & \hspace*{-1.23cm}   2.404 & \hspace*{-1.23cm} $\pm   0.122$ & \hspace*{-1.23cm} $^{+   0.055}_{-  0.055}$ & \hspace*{-1.23cm} $^{+   0.034}_{-  0.035}$ \\
 \hline
  22.00 --   28.00 & \hspace*{-1.23cm}   2.127 & \hspace*{-1.23cm} $\pm   0.109$ & \hspace*{-1.23cm} $^{+   0.048}_{-  0.041}$ & \hspace*{-1.23cm} $^{+   0.057}_{-  0.053}$ \\
 \hline
  28.00 --   35.00 & \hspace*{-1.23cm}   1.272 & \hspace*{-1.23cm} $\pm   0.077$ & \hspace*{-1.23cm} $^{+   0.060}_{-  0.038}$ & \hspace*{-1.23cm} $^{+   0.046}_{-  0.042}$ \\
 \hline
  35.00 --   45.00 &   7.494$\; \cdot \; 10^{-1}$ & $(\pm   0.497)\cdot 10^{-1}$ & $(^{+   0.090}_{-  0.309})\cdot 10^{-1}$ & $(^{+   0.256}_{-  0.246})\cdot 10^{-1}$ \\
 \hline
  45.00 --   60.00 &   3.430$\; \cdot \; 10^{-1}$ & $(\pm   0.284)\cdot 10^{-1}$ & $(^{+   0.100}_{-  0.484})\cdot 10^{-1}$ & $(^{+   0.270}_{-  0.195})\cdot 10^{-1}$ \\
 \hline
  60.00 --   80.00 &   9.605$\; \cdot \; 10^{-2}$ & $(\pm   1.272)\cdot 10^{-2}$ & $(^{+   0.326}_{-  0.419})\cdot 10^{-2}$ & $(^{+   0.619}_{-  0.709})\cdot 10^{-2}$ \\
 \hline
  80.00 --  120.00 &   1.578$\; \cdot \; 10^{-2}$ & $(\pm   0.353)\cdot 10^{-2}$ & $(^{+   0.122}_{-  0.219})\cdot 10^{-2}$ & $(^{+   0.117}_{-  0.145})\cdot 10^{-2}$ \\
 \hline
 \end{tabular}
 \end{center}
 \caption{\label{table4} The differential dijet cross section $d\sigma_{2+1}/dM_{jj}$.
 Other details are as described in the caption to Table~\ref{table1}.}
 \vspace{0.3cm}
 \end{table}

 \begin{table}
 \begin{center}
 \hspace*{-0.4cm}
 \begin{tabular}{|c|c|c|c|c|}
 \hline
 \hline
 $E_{T,B}^{jet,1}$ range [GeV]&                                             
 $d\sigma_{2+1}/dE_{T,B}^{jet,1}$~[pb/GeV] &                           
 $\Delta_{\rm stat}$ &
 $\Delta_{\rm syst}$ &
 $\Delta_{\rm ES}$ \\
 \hline
 \hline
   5.00 --    8.00 & \hspace*{-1.23cm}   2.956 & \hspace*{-1.23cm} $\pm   0.175$ & \hspace*{-1.23cm} $^{+   0.154}_{-  0.243}$ & \hspace*{-1.23cm} $^{+   0.072}_{-  0.081}$ \\
 \hline
   8.00 --   12.00 & \hspace*{-1.23cm}   5.977 & \hspace*{-1.23cm} $\pm   0.238$ & \hspace*{-1.23cm} $^{+   0.105}_{-  0.201}$ & \hspace*{-1.23cm} $^{+   0.094}_{-  0.090}$ \\
 \hline
  12.00 --   16.00 & \hspace*{-1.23cm}   2.736 & \hspace*{-1.23cm} $\pm   0.154$ & \hspace*{-1.23cm} $^{+   0.039}_{-  0.033}$ & \hspace*{-1.23cm} $^{+   0.083}_{-  0.081}$ \\
 \hline
  16.00 --   20.00 & \hspace*{-1.23cm}   1.327 & \hspace*{-1.23cm} $\pm   0.103$ & \hspace*{-1.23cm} $^{+   0.055}_{-  0.030}$ & \hspace*{-1.23cm} $^{+   0.072}_{-  0.026}$ \\
 \hline
  20.00 --   24.00 &   8.165$\; \cdot \; 10^{-1}$ & $(\pm   0.801)\cdot 10^{-1}$ & $(^{+   0.096}_{-  0.397})\cdot 10^{-1}$ & $(^{+   0.245}_{-  0.566})\cdot 10^{-1}$ \\
 \hline
  24.00 --   32.00 &   2.848$\; \cdot \; 10^{-1}$ & $(\pm   0.333)\cdot 10^{-1}$ & $(^{+   0.077}_{-  0.355})\cdot 10^{-1}$ & $(^{+   0.236}_{-  0.183})\cdot 10^{-1}$ \\
 \hline
  32.00 --   50.00 &   5.960$\; \cdot \; 10^{-2}$ & $(\pm   0.879)\cdot 10^{-2}$ & $(^{+   0.997}_{-  0.449})\cdot 10^{-2}$ & $(^{+   0.572}_{-  0.497})\cdot 10^{-2}$ \\
 \hline
  50.00 --   80.00 &   3.991$\; \cdot \; 10^{-3}$ & $(\pm   1.203)\cdot 10^{-3}$ & $(^{+   1.086}_{-  0.902})\cdot 10^{-3}$ & $(^{+   0.548}_{-  0.472})\cdot 10^{-3}$ \\
 \hline
 \end{tabular}
 \end{center}
 \caption{\label{table5} The differential dijet cross section $d\sigma_{2+1}/dE_{T,B}^{jet,1}$.
 Other details are as described in the caption to Table~\ref{table1}.}
 \end{table}

\newpage
\clearpage
 \begin{table}
 \begin{center}
 \hspace*{0.2cm}
 \begin{tabular}{|c|c|c|c|c|}
 \hline
 \hline
 $\eta_B^{jet,1}$ range &                                              
 $d\sigma_{2+1}/d\eta_B^{jet,1}$~[pb] &                                
 $\Delta_{\rm stat}$ &
 $\Delta_{\rm syst}$ &
 $\Delta_{\rm ES}$ \\
 \hline
 \hline
-1.00 -- -0.50 & \hspace*{-1.00cm}   3.051 & \hspace*{-1.00cm} $\pm   0.419$ & \hspace*{-1.00cm} $^{+   0.360}_{-  0.277}$ & \hspace*{-1.00cm} $^{+   0.155}_{-  0.141}$ \\
 \hline
-0.50 --  0.00 &   1.170$\; \cdot \; 10^{ 1}$ & $(\pm   0.084)\cdot 10^{ 1}$ & $(^{+   0.025}_{-  0.158})\cdot 10^{ 1}$ & $(^{+   0.060}_{-  0.051})\cdot 10^{ 1}$ \\
 \hline
 0.00 --  0.50 &   3.260$\; \cdot \; 10^{ 1}$ & $(\pm   0.154)\cdot 10^{ 1}$ & $(^{+   0.116}_{-  0.105})\cdot 10^{ 1}$ & $(^{+   0.102}_{-  0.091})\cdot 10^{ 1}$ \\
 \hline
 0.50 --  1.00 &   3.204$\; \cdot \; 10^{ 1}$ & $(\pm   0.151)\cdot 10^{ 1}$ & $(^{+   0.084}_{-  0.147})\cdot 10^{ 1}$ & $(^{+   0.068}_{-  0.075})\cdot 10^{ 1}$ \\
 \hline
 1.00 --  1.50 &   2.054$\; \cdot \; 10^{ 1}$ & $(\pm   0.119)\cdot 10^{ 1}$ & $(^{+   0.083}_{-  0.104})\cdot 10^{ 1}$ & $(^{+   0.046}_{-  0.049})\cdot 10^{ 1}$ \\
 \hline
 1.50 --  2.00 & \hspace*{-1.00cm}   9.035 & \hspace*{-1.00cm} $\pm   0.748$ & \hspace*{-1.00cm} $^{+   0.570}_{-  0.742}$ & \hspace*{-1.00cm} $^{+   0.229}_{-  0.194}$ \\
 \hline
 2.00 --  2.50 & \hspace*{-1.00cm}   2.561 & \hspace*{-1.00cm} $\pm   0.410$ & \hspace*{-1.00cm} $^{+   0.491}_{-  0.324}$ & \hspace*{-1.00cm} $^{+   0.061}_{-  0.058}$ \\
 \hline
 \end{tabular}
 \end{center}
 \caption{\label{table6} The differential dijet cross section $d\sigma_{2+1}/d\eta_B^{jet,1}$.
 Other details are as described in the caption to Table~\ref{table1}.}
 \vspace{0.3cm}
 \end{table}

 \begin{table}
 \begin{center}
 \hspace*{-0.5cm}
 \begin{tabular}{|c|c|c|c|c|}
 \hline
 \hline
 $E_{T,B}^{jet,2}$ range [GeV]&                                             
 $d\sigma_{2+1}/dE_{T,B}^{jet,2}$~[pb/GeV] &                           
 $\Delta_{\rm stat}$ &
 $\Delta_{\rm syst}$ &
 $\Delta_{\rm ES}$ \\
 \hline
 \hline
   5.00 --    8.00 & \hspace*{-1.23cm}   2.775 & \hspace*{-1.23cm} $\pm   0.168$ & \hspace*{-1.23cm} $^{+   0.098}_{-  0.235}$ & \hspace*{-1.23cm} $^{+   0.074}_{-  0.082}$ \\
 \hline
   8.00 --   12.00 & \hspace*{-1.23cm}   5.971 & \hspace*{-1.23cm} $\pm   0.240$ & \hspace*{-1.23cm} $^{+   0.142}_{-  0.154}$ & \hspace*{-1.23cm} $^{+   0.103}_{-  0.092}$ \\
 \hline
  12.00 --   16.00 & \hspace*{-1.23cm}   2.762 & \hspace*{-1.23cm} $\pm   0.152$ & \hspace*{-1.23cm} $^{+   0.074}_{-  0.071}$ & \hspace*{-1.23cm} $^{+   0.088}_{-  0.076}$ \\
 \hline
  16.00 --   20.00 & \hspace*{-1.23cm}   1.403 & \hspace*{-1.23cm} $\pm   0.105$ & \hspace*{-1.23cm} $^{+   0.011}_{-  0.034}$ & \hspace*{-1.23cm} $^{+   0.062}_{-  0.050}$ \\
 \hline
  20.00 --   24.00 &   7.437$\; \cdot \; 10^{-1}$ & $(\pm   0.763)\cdot 10^{-1}$ & $(^{+   0.144}_{-  1.274})\cdot 10^{-1}$ & $(^{+   0.397}_{-  0.406})\cdot 10^{-1}$ \\
 \hline
  24.00 --   32.00 &   2.995$\; \cdot \; 10^{-1}$ & $(\pm   0.339)\cdot 10^{-1}$ & $(^{+   0.070}_{-  0.507})\cdot 10^{-1}$ & $(^{+   0.182}_{-  0.198})\cdot 10^{-1}$ \\
 \hline
  32.00 --   50.00 &   6.834$\; \cdot \; 10^{-2}$ & $(\pm   0.966)\cdot 10^{-2}$ & $(^{+   0.641}_{-  0.239})\cdot 10^{-2}$ & $(^{+   0.551}_{-  0.452})\cdot 10^{-2}$ \\
 \hline
  50.00 --   80.00 &   3.189$\; \cdot \; 10^{-3}$ & $(\pm   1.205)\cdot 10^{-3}$ & $(^{+   0.981}_{-  0.621})\cdot 10^{-3}$ & $(^{+   0.203}_{-  0.251})\cdot 10^{-3}$ \\
 \hline
 \end{tabular}
 \end{center}
 \caption{\label{table7} The differential dijet cross section $d\sigma_{2+1}/dE_{T,B}^{jet,2}$.
 Other details are as described in the caption to Table~\ref{table1}.}
 \vspace{0.3cm}
 \end{table}

 \begin{table}
 \begin{center}
 \hspace*{0.2cm}
 \begin{tabular}{|c|c|c|c|c|}
 \hline
 \hline
 $\eta_B^{jet,2}$ range &                                              
 $d\sigma_{2+1}/d\eta_B^{jet,2}$~[pb] &                                
 $\Delta_{\rm stat}$ &
 $\Delta_{\rm syst}$ &
 $\Delta_{\rm ES}$ \\
 \hline
 \hline
-2.00 -- -1.50 & \hspace*{-1.00cm}   2.507 & \hspace*{-1.00cm} $\pm   0.458$ & \hspace*{-1.00cm} $^{+   0.219}_{-  0.670}$ & \hspace*{-1.00cm} $^{+   0.202}_{-  0.190}$ \\
 \hline
-1.50 -- -1.00 &   1.054$\; \cdot \; 10^{ 1}$ & $(\pm   0.084)\cdot 10^{ 1}$ & $(^{+   0.043}_{-  0.084})\cdot 10^{ 1}$ & $(^{+   0.069}_{-  0.067})\cdot 10^{ 1}$ \\
 \hline
-1.00 -- -0.50 &   3.328$\; \cdot \; 10^{ 1}$ & $(\pm   0.151)\cdot 10^{ 1}$ & $(^{+   0.080}_{-  0.170})\cdot 10^{ 1}$ & $(^{+   0.154}_{-  0.134})\cdot 10^{ 1}$ \\
 \hline
-0.50 --  0.00 &   3.735$\; \cdot \; 10^{ 1}$ & $(\pm   0.163)\cdot 10^{ 1}$ & $(^{+   0.067}_{-  0.130})\cdot 10^{ 1}$ & $(^{+   0.073}_{-  0.066})\cdot 10^{ 1}$ \\
 \hline
 0.00 --  0.50 &   1.885$\; \cdot \; 10^{ 1}$ & $(\pm   0.112)\cdot 10^{ 1}$ & $(^{+   0.062}_{-  0.066})\cdot 10^{ 1}$ & $(^{+   0.012}_{-  0.014})\cdot 10^{ 1}$ \\
 \hline
 0.50 --  1.00 & \hspace*{-1.00cm}   6.868 & \hspace*{-1.00cm} $\pm   0.627$ & \hspace*{-1.00cm} $^{+   0.504}_{-  0.252}$ & \hspace*{-1.00cm} $^{+   0.070}_{-  0.080}$ \\
 \hline
 1.00 --  1.50 & \hspace*{-1.00cm}   1.914 & \hspace*{-1.00cm} $\pm   0.349$ & \hspace*{-1.00cm} $^{+   0.288}_{-  0.273}$ & \hspace*{-1.00cm} $^{+   0.066}_{-  0.057}$ \\
 \hline
 \end{tabular}
 \end{center}
 \caption{\label{table8} The differential dijet cross section $d\sigma_{2+1}/d\eta_B^{jet,2}$.
 Other details are as described in the caption to Table~\ref{table1}.}
 \end{table}

\newpage
\clearpage
 \begin{table}
 \begin{center}
 \begin{tabular}{|c|c|c|c|}
 \hline
 \hline
 $Q^2$ range [GeV$^2$]&                                                         
 $d\sigma_{\rm tot}/dQ^2$~[pb/GeV$^2$] &                                   
 $\Delta_{\rm stat}$ &
 $\Delta_{\rm syst}$ \\
 \hline
 \hline
     470. --      800. & \hspace*{-1.23cm}   1.217 & \hspace*{-1.23cm} $\pm   0.011$ & \hspace*{-1.23cm} $^{+   0.015}_{-  0.006}$ \\
 \hline
     800. --     1500. &   2.800$\; \cdot \; 10^{-1}$ & $(\pm   0.033)\cdot 10^{-1}$ & $(^{+   0.023}_{-  0.019})\cdot 10^{-1}$ \\
 \hline
    1500. --     2500. &   6.509$\; \cdot \; 10^{-2}$ & $(\pm   0.130)\cdot 10^{-2}$ & $(^{+   0.103}_{-  0.090})\cdot 10^{-2}$ \\
 \hline
    2500. --     5000. &   1.230$\; \cdot \; 10^{-2}$ & $(\pm   0.036)\cdot 10^{-2}$ & $(^{+   0.022}_{-  0.020})\cdot 10^{-2}$ \\
 \hline
    5000. --    20000. &   7.378$\; \cdot \; 10^{-4}$ & $(\pm   0.400)\cdot 10^{-4}$ & $(^{+   0.051}_{-  0.163})\cdot 10^{-4}$ \\
 \hline
 \end{tabular}
 \end{center}
 \caption{\label{table9} The differential inclusive cross section $d\sigma_{\rm tot}/dQ^2$.
 Other details are as described in the caption to Table~\ref{table1}.}
 \end{table}

 \begin{table}
 \begin{center}
 \begin{tabular}{|c|c|c|c|c|}
 \hline
 \hline
 $Q^2$ range [GeV$^2$]&                                                         
 $d\sigma_{2+1}/dQ^2$~[pb/GeV$^2$] &                                   
 $\Delta_{\rm stat}$ &
 $\Delta_{\rm syst}$ &
 $\Delta_{\rm ES}$ \\
 \hline
 \hline
     470. --      800. &   8.061$\; \cdot \; 10^{-2}$ & $(\pm   0.306)\cdot 10^{-2}$ & $(^{+   0.212}_{-  0.294})\cdot 10^{-2}$ & $(^{+   0.287}_{-  0.256})\cdot 10^{-2}$ \\
 \hline
     800. --     1500. &   2.415$\; \cdot \; 10^{-2}$ & $(\pm   0.105)\cdot 10^{-2}$ & $(^{+   0.039}_{-  0.061})\cdot 10^{-2}$ & $(^{+   0.071}_{-  0.072})\cdot 10^{-2}$ \\
 \hline
    1500. --     2500. &   6.802$\; \cdot \; 10^{-3}$ & $(\pm   0.453)\cdot 10^{-3}$ & $(^{+   0.300}_{-  0.343})\cdot 10^{-3}$ & $(^{+   0.159}_{-  0.160})\cdot 10^{-3}$ \\
 \hline
    2500. --     5000. &   1.637$\; \cdot \; 10^{-3}$ & $(\pm   0.139)\cdot 10^{-3}$ & $(^{+   0.046}_{-  0.124})\cdot 10^{-3}$ & $(^{+   0.033}_{-  0.028})\cdot 10^{-3}$ \\
 \hline
    5000. --    20000. &   1.246$\; \cdot \; 10^{-4}$ & $(\pm   0.170)\cdot 10^{-4}$ & $(^{+   0.006}_{-  0.139})\cdot 10^{-4}$ & $(^{+   0.013}_{-  0.018})\cdot 10^{-4}$ \\
 \hline
 \end{tabular}
 \end{center}
 \caption{\label{table10} The differential dijet cross section $d\sigma_{2+1}/dQ^2$.
 Other details are as described in the caption to Table~\ref{table1}.}
 \end{table}

 \begin{table}
 \begin{center}
 \begin{tabular}{|c|c|c|c|c|}
 \hline
 \hline
 $Q^2$ range [GeV$^2$]&                                                         
 $R_{2+1}$ &                                                           
 $\Delta_{\rm stat}$ &
 $\Delta_{\rm syst}$ &
 $\Delta_{\rm ES}$ \\
 \hline
 \hline
     470. --      800. &   6.625$\; \cdot \; 10^{-2}$ & $(\pm   0.245)\cdot 10^{-2}$ & $(^{+   0.168}_{-  0.278})\cdot 10^{-2}$ & $(^{+   0.236}_{-  0.210})\cdot 10^{-2}$ \\
 \hline
     800. --     1500. &   8.627$\; \cdot \; 10^{-2}$ & $(\pm   0.362)\cdot 10^{-2}$ & $(^{+   0.144}_{-  0.267})\cdot 10^{-2}$ & $(^{+   0.255}_{-  0.256})\cdot 10^{-2}$ \\
 \hline
    1500. --     2500. &   1.045$\; \cdot \; 10^{-1}$ & $(\pm   0.066)\cdot 10^{-1}$ & $(^{+   0.043}_{-  0.057})\cdot 10^{-1}$ & $(^{+   0.024}_{-  0.025})\cdot 10^{-1}$ \\
 \hline
    2500. --     5000. &   1.331$\; \cdot \; 10^{-1}$ & $(\pm   0.106)\cdot 10^{-1}$ & $(^{+   0.039}_{-  0.119})\cdot 10^{-1}$ & $(^{+   0.027}_{-  0.023})\cdot 10^{-1}$ \\
 \hline
    5000. --    20000. &   1.688$\; \cdot \; 10^{-1}$ & $(\pm   0.211)\cdot 10^{-1}$ & $(^{+   0.008}_{-  0.185})\cdot 10^{-1}$ & $(^{+   0.017}_{-  0.024})\cdot 10^{-1}$ \\
 \hline
 \end{tabular}
 \end{center}
 \caption{\label{table11} The dijet fraction $R_{2+1}(Q^2)$.
 Other details are as described in the caption to Table~\ref{table1}.}
 \end{table}

%
%
\newpage
\begin{landscape}
\clearpage
%
\catcode`\@=11 
\renewcommand{\@makecaption}[2]%
  {\def\baselinestretch{0.95}%
   \vspace{10.pt}
   \setlength{\@captionwidth}{23cm}
   \addtolength{\@captionwidth}{-\@captionmargin}
   \sbox{\tmpbox}{{\bf #1:}{\rm #2}}%
   \ifthenelse{\lengthtest{\wd\tmpbox > \@captionwidth}}%
   {\centerline{\parbox[t]{\@captionwidth}%
   {\tolerance=2000\normalsize%
    {\bf #1:}\hspace{\@captionitemtextsep}{\rm #2}}}}%
   {\centerline{{\bf #1:}\kern1.em{\rm #2}}}}
\catcode`\@=12 
 \begin{table} [!hb]
 \begin{center}
 \begin{tabular}{|c|c|c|c|c|c|c|c|}
 \hline
 \hline
 $z_{p,1}$ range &                                                         
 $<z_{p,1}>$ &                                                             
 $d\sigma_{2+1}^{NLO}/dz_{p,1}$~[pb] &                                     
 $\Delta_{\mu_r}$ &
 $\Delta_{\alpha_s}$ &
 $\Delta^{\rm exp}_{pdf}$ &
 $\Delta^{\rm theo}_{pdf}$ &
 $C_{\rm had}$ $\pm$ $\Delta C_{\rm had}$ \\
 \hline
 \hline
 0.00 --  0.05 &  0.04 &   1.854$\; \cdot \; 10^{ 1}$ & $(^{+   0.256}_{-  0.197})\cdot 10^{ 1}$ & $(^{+   0.147}_{-  0.143})\cdot 10^{ 1}$ & $(\pm   0.105)\cdot 10^{ 1}$ & $(^{+   0.023}_{-  0.027})\cdot 10^{ 1}$ &    1.38$\pm$    0.04 \\
 \hline
 0.05 --  0.15 &  0.10 &   1.172$\; \cdot \; 10^{ 2}$ & $(^{+   0.100}_{-  0.088})\cdot 10^{ 2}$ & $(^{+   0.078}_{-  0.082})\cdot 10^{ 2}$ & $(\pm   0.060)\cdot 10^{ 2}$ & $(^{+   0.014}_{-  0.014})\cdot 10^{ 2}$ &    1.10$\pm$    0.01 \\
 \hline
 0.15 --  0.25 &  0.20 &   1.167$\; \cdot \; 10^{ 2}$ & $(^{+   0.056}_{-  0.062})\cdot 10^{ 2}$ & $(^{+   0.070}_{-  0.075})\cdot 10^{ 2}$ & $(\pm   0.067)\cdot 10^{ 2}$ & $(^{+   0.015}_{-  0.014})\cdot 10^{ 2}$ &    1.08$\pm$    0.04 \\
 \hline
 0.25 --  0.35 &  0.30 &   9.100$\; \cdot \; 10^{ 1}$ & $(^{+   0.351}_{-  0.432})\cdot 10^{ 1}$ & $(^{+   0.542}_{-  0.577})\cdot 10^{ 1}$ & $(\pm   0.529)\cdot 10^{ 1}$ & $(^{+   0.120}_{-  0.118})\cdot 10^{ 1}$ &    1.07$\pm$    0.01 \\
 \hline
 0.35 --  0.45 &  0.40 &   7.481$\; \cdot \; 10^{ 1}$ & $(^{+   0.280}_{-  0.349})\cdot 10^{ 1}$ & $(^{+   0.454}_{-  0.477})\cdot 10^{ 1}$ & $(\pm   0.421)\cdot 10^{ 1}$ & $(^{+   0.100}_{-  0.097})\cdot 10^{ 1}$ &    1.07$\pm$    0.02 \\
 \hline
 0.45 --  0.55 &  0.50 &   6.667$\; \cdot \; 10^{ 1}$ & $(^{+   0.275}_{-  0.327})\cdot 10^{ 1}$ & $(^{+   0.403}_{-  0.425})\cdot 10^{ 1}$ & $(\pm   0.361)\cdot 10^{ 1}$ & $(^{+   0.089}_{-  0.085})\cdot 10^{ 1}$ &    1.09$\pm$    0.02 \\
 \hline
 0.55 --  0.65 &  0.60 &   5.711$\; \cdot \; 10^{ 1}$ & $(^{+   0.185}_{-  0.250})\cdot 10^{ 1}$ & $(^{+   0.339}_{-  0.356})\cdot 10^{ 1}$ & $(\pm   0.349)\cdot 10^{ 1}$ & $(^{+   0.074}_{-  0.070})\cdot 10^{ 1}$ &    1.09$\pm$    0.01 \\
 \hline
 0.65 --  0.75 &  0.70 &   4.858$\; \cdot \; 10^{ 1}$ & $(^{+   0.219}_{-  0.248})\cdot 10^{ 1}$ & $(^{+   0.294}_{-  0.308})\cdot 10^{ 1}$ & $(\pm   0.309)\cdot 10^{ 1}$ & $(^{+   0.061}_{-  0.058})\cdot 10^{ 1}$ &    1.15$\pm$    0.03 \\
 \hline
 0.75 --  0.85 &  0.79 &   2.776$\; \cdot \; 10^{ 1}$ & $(^{+   0.215}_{-  0.194})\cdot 10^{ 1}$ & $(^{+   0.202}_{-  0.193})\cdot 10^{ 1}$ & $(\pm   0.168)\cdot 10^{ 1}$ & $(^{+   0.037}_{-  0.036})\cdot 10^{ 1}$ &    1.28$\pm$    0.00 \\
 \hline
 0.85 --  0.95 &  0.88 & \hspace*{-1.00cm}   4.466 & \hspace*{-1.00cm} $^{+   0.573}_{-  0.445}$ & \hspace*{-1.00cm} $^{+   0.377}_{-  0.318}$ & \hspace*{-1.00cm} $\pm   0.265$ & \hspace*{-1.00cm} $^{+   0.049}_{-  0.072}$ &    1.27$\pm$    0.13 \\
 \hline
 \end{tabular}
 \end{center}
 \caption{\label{table13}
  The QCD predictions for the differential dijet cross section as a
  function of $z_{p,1}$. For each bin in $z_{p,1}$ the following quantities are given:
  the weighted mean value $<z_{p,1}>$, the pure NLO QCD cross section, the uncertainty due 
  to the renormalization scale, $\Delta_{\mu_r}$, the uncertainty due to $\asz$,
  $\Delta_{\alpha_s}$, the uncertainty due to the proton PDFs (experimental),
  $\Delta^{\rm exp}_{pdf}$, the uncertainty due to the proton PDFs (theoretical),
  $\Delta^{\rm theo}_{pdf}$, and the hadronisation correction, $C_{\rm had}$, 
  with its associated uncertainty, $\Delta C_{\rm had}$.}
 \end{table}
 
 \begin{table} [!hb]
 \begin{center}
 \hspace*{-0.4cm}
 \begin{tabular}{|c|c|c|c|c|c|c|c|}
 \hline
 \hline
 $\log_{10}(x_{Bj})$ range &                                           
 $<\log_{10}(x_{Bj})>$ &                                               
 $d\sigma_{2+1}^{NLO}/d\log_{10}(x_{Bj})$~[pb] &                       
 $\Delta_{\mu_r}$ &
 $\Delta_{\alpha_s}$ &
 $\Delta^{\rm exp}_{pdf}$ &
 $\Delta^{\rm theo}_{pdf}$ &
 $C_{\rm had}$ $\pm$ $\Delta C_{\rm had}$ \\
 \hline
 \hline
-2.2 -- -2.0 & -2.08 &   2.509$\; \cdot \; 10^{ 1}$ & $(^{+   0.184}_{-  0.173})\cdot 10^{ 1}$ & $(^{+   0.121}_{-  0.157})\cdot 10^{ 1}$ & $(\pm   0.140)\cdot 10^{ 1}$ & $(^{+   0.036}_{-  0.040})\cdot 10^{ 1}$ &    1.04$\pm$    0.02 \\
 \hline
-2.0 -- -1.8 & -1.90 &   5.025$\; \cdot \; 10^{ 1}$ & $(^{+   0.285}_{-  0.297})\cdot 10^{ 1}$ & $(^{+   0.246}_{-  0.311})\cdot 10^{ 1}$ & $(\pm   0.248)\cdot 10^{ 1}$ & $(^{+   0.064}_{-  0.072})\cdot 10^{ 1}$ &    1.06$\pm$    0.01 \\
 \hline
-1.8 -- -1.6 & -1.70 &   6.477$\; \cdot \; 10^{ 1}$ & $(^{+   0.336}_{-  0.361})\cdot 10^{ 1}$ & $(^{+   0.363}_{-  0.419})\cdot 10^{ 1}$ & $(\pm   0.256)\cdot 10^{ 1}$ & $(^{+   0.076}_{-  0.078})\cdot 10^{ 1}$ &    1.08$\pm$    0.02 \\
 \hline
-1.6 -- -1.4 & -1.50 &   6.403$\; \cdot \; 10^{ 1}$ & $(^{+   0.311}_{-  0.341})\cdot 10^{ 1}$ & $(^{+   0.423}_{-  0.436})\cdot 10^{ 1}$ & $(\pm   0.205)\cdot 10^{ 1}$ & $(^{+   0.079}_{-  0.068})\cdot 10^{ 1}$ &    1.10$\pm$    0.01 \\
 \hline
-1.4 -- -1.2 & -1.31 &   5.003$\; \cdot \; 10^{ 1}$ & $(^{+   0.247}_{-  0.266})\cdot 10^{ 1}$ & $(^{+   0.377}_{-  0.353})\cdot 10^{ 1}$ & $(\pm   0.165)\cdot 10^{ 1}$ & $(^{+   0.072}_{-  0.059})\cdot 10^{ 1}$ &    1.14$\pm$    0.01 \\
 \hline
-1.2 -- -1.0 & -1.11 &   2.935$\; \cdot \; 10^{ 1}$ & $(^{+   0.148}_{-  0.155})\cdot 10^{ 1}$ & $(^{+   0.231}_{-  0.201})\cdot 10^{ 1}$ & $(\pm   0.122)\cdot 10^{ 1}$ & $(^{+   0.044}_{-  0.037})\cdot 10^{ 1}$ &    1.15$\pm$    0.02 \\
 \hline
-1.0 -- -0.8 & -0.91 &   1.391$\; \cdot \; 10^{ 1}$ & $(^{+   0.070}_{-  0.071})\cdot 10^{ 1}$ & $(^{+   0.100}_{-  0.083})\cdot 10^{ 1}$ & $(\pm   0.064)\cdot 10^{ 1}$ & $(^{+   0.017}_{-  0.017})\cdot 10^{ 1}$ &    1.13$\pm$    0.01 \\
 \hline
-0.8 -- -0.6 & -0.72 & \hspace*{-1.00cm}   5.604 & \hspace*{-1.00cm} $^{+   0.283}_{-  0.282}$ & \hspace*{-1.00cm} $^{+   0.322}_{-  0.263}$ & \hspace*{-1.00cm} $\pm   0.244$ & \hspace*{-1.00cm} $^{+   0.041}_{-  0.069}$ &    1.11$\pm$    0.01 \\
 \hline
-0.6 -- -0.4 & -0.52 & \hspace*{-1.00cm}   1.650 & \hspace*{-1.00cm} $^{+   0.088}_{-  0.082}$ & \hspace*{-1.00cm} $^{+   0.061}_{-  0.051}$ & \hspace*{-1.00cm} $\pm   0.062$ & \hspace*{-1.00cm} $^{+   0.000}_{-  0.023}$ &    1.14$\pm$    0.05 \\
 \hline
 \end{tabular}
 \end{center}
 \caption{\label{table14} The QCD predictions for the differential dijet cross section as a
  function of $\log_{10}(x_{Bj})$. Other details are as described in the caption to
  Table~\ref{table13}.}
 \end{table}

 \begin{table} [!hb]
 \begin{center}
 \begin{tabular}{|c|c|c|c|c|c|c|c|}
 \hline
 \hline
 $\log_{10}(\xi)$ range &                                              
 $<\log_{10}(\xi)>$ &                                                  
 $d\sigma_{2+1}^{NLO}/d\log_{10}(\xi)$~[pb] &                          
 $\Delta_{\mu_r}$ &
 $\Delta_{\alpha_s}$ &
 $\Delta^{\rm exp}_{pdf}$ &
 $\Delta^{\rm theo}_{pdf}$ &
 $C_{\rm had}$ $\pm$ $\Delta C_{\rm had}$ \\
 \hline
 \hline
  -2.1875 --   -1.8750 & -1.94 & \hspace*{-1.00cm}   2.825 & \hspace*{-1.00cm} $^{+   0.737}_{-  0.503}$ & \hspace*{-1.00cm} $^{+   0.233}_{-  0.258}$ & \hspace*{-1.00cm} $\pm   0.166$ & \hspace*{-1.00cm} $^{+   0.047}_{-  0.051}$ &    1.18$\pm$    0.05 \\
 \hline
  -1.8750 --   -1.5625 & -1.68 &   2.951$\; \cdot \; 10^{ 1}$ & $(^{+   0.363}_{-  0.290})\cdot 10^{ 1}$ & $(^{+   0.167}_{-  0.205})\cdot 10^{ 1}$ & $(\pm   0.180)\cdot 10^{ 1}$ & $(^{+   0.043}_{-  0.048})\cdot 10^{ 1}$ &    1.09$\pm$    0.01 \\
 \hline
  -1.5625 --   -1.2500 & -1.40 &   6.736$\; \cdot \; 10^{ 1}$ & $(^{+   0.416}_{-  0.416})\cdot 10^{ 1}$ & $(^{+   0.372}_{-  0.434})\cdot 10^{ 1}$ & $(\pm   0.318)\cdot 10^{ 1}$ & $(^{+   0.075}_{-  0.076})\cdot 10^{ 1}$ &    1.06$\pm$    0.01 \\
 \hline
  -1.2500 --   -0.9375 & -1.11 &   6.746$\; \cdot \; 10^{ 1}$ & $(^{+   0.198}_{-  0.282})\cdot 10^{ 1}$ & $(^{+   0.453}_{-  0.448})\cdot 10^{ 1}$ & $(\pm   0.273)\cdot 10^{ 1}$ & $(^{+   0.085}_{-  0.069})\cdot 10^{ 1}$ &    1.11$\pm$    0.01 \\
 \hline
  -0.9375 --   -0.6250 & -0.82 &   2.569$\; \cdot \; 10^{ 1}$ & $(^{+   0.002}_{-  0.059})\cdot 10^{ 1}$ & $(^{+   0.190}_{-  0.157})\cdot 10^{ 1}$ & $(\pm   0.162)\cdot 10^{ 1}$ & $(^{+   0.041}_{-  0.040})\cdot 10^{ 1}$ &    1.14$\pm$    0.02 \\
 \hline
  -0.6250 --   -0.3125 & -0.53 & \hspace*{-1.00cm}   3.378 & \hspace*{-1.00cm} $^{+   0.000}_{-  0.059}$ & \hspace*{-1.00cm} $^{+   0.177}_{-  0.122}$ & \hspace*{-1.00cm} $\pm   0.220$ & \hspace*{-1.00cm} $^{+   0.027}_{-  0.067}$ &    1.11$\pm$    0.06 \\
 \hline
 \end{tabular}
 \end{center}
 \caption{\label{table15} The QCD predictions for the differential dijet cross section as a
  function of $\log_{10}(\xi)$. Other details are as described in the caption to
  Table~\ref{table13}.}
 \end{table}

 \begin{table} [!hb]
 \begin{center}
 \begin{tabular}{|c|c|c|c|c|c|c|c|}
 \hline
 \hline
 $M_{jj}$ range [GeV]&                                                      
 $<M_{jj}>$ &                                                          
 $d\sigma_{2+1}^{NLO}/dM_{jj}$~[pb/GeV] &                              
 $\Delta_{\mu_r}$ &
 $\Delta_{\alpha_s}$ &
 $\Delta^{\rm exp}_{pdf}$ &
 $\Delta^{\rm theo}_{pdf}$ &
 $C_{\rm had}$ $\pm$ $\Delta C_{\rm had}$ \\
 \hline
 \hline
  10. --   16. &   14.65 &   6.351$\; \cdot \; 10^{-1}$ & $(^{+   2.017}_{-  1.355})\cdot 10^{-1}$ & $(^{+   0.751}_{-  0.684})\cdot 10^{-1}$ & $(\pm   0.190)\cdot 10^{-1}$ & $(^{+   0.099}_{-  0.112})\cdot 10^{-1}$ &    1.16$\pm$    0.02 \\
 \hline
  16. --   22. &   19.03 & \hspace*{-1.23cm}   2.914 & \hspace*{-1.23cm} $^{+   0.346}_{-  0.278}$ & \hspace*{-1.23cm} $^{+   0.208}_{-  0.216}$ & \hspace*{-1.23cm} $\pm   0.149$ & \hspace*{-1.23cm} $^{+   0.038}_{-  0.036}$ &    1.14$\pm$    0.03 \\
 \hline
  22. --   28. &   24.98 & \hspace*{-1.23cm}   2.331 & \hspace*{-1.23cm} $^{+   0.079}_{-  0.104}$ & \hspace*{-1.23cm} $^{+   0.126}_{-  0.141}$ & \hspace*{-1.23cm} $\pm   0.161$ & \hspace*{-1.23cm} $^{+   0.028}_{-  0.026}$ &    1.13$\pm$    0.02 \\
 \hline
  28. --   35. &   31.32 & \hspace*{-1.23cm}   1.487 & \hspace*{-1.23cm} $^{+   0.000}_{-  0.035}$ & \hspace*{-1.23cm} $^{+   0.073}_{-  0.083}$ & \hspace*{-1.23cm} $\pm   0.112$ & \hspace*{-1.23cm} $^{+   0.017}_{-  0.016}$ &    1.05$\pm$    0.01 \\
 \hline
  35. --   45. &   39.42 &   8.119$\; \cdot \; 10^{-1}$ & $(^{+   0.000}_{-  0.174})\cdot 10^{-1}$ & $(^{+   0.403}_{-  0.446})\cdot 10^{-1}$ & $(\pm   0.567)\cdot 10^{-1}$ & $(^{+   0.095}_{-  0.091})\cdot 10^{-1}$ &    1.04$\pm$    0.02 \\
 \hline
  45. --   60. &   51.11 &   3.227$\; \cdot \; 10^{-1}$ & $(^{+   0.000}_{-  0.111})\cdot 10^{-1}$ & $(^{+   0.174}_{-  0.179})\cdot 10^{-1}$ & $(\pm   0.223)\cdot 10^{-1}$ & $(^{+   0.042}_{-  0.041})\cdot 10^{-1}$ &    1.00$\pm$    0.02 \\
 \hline
  60. --   80. &   67.73 &   9.154$\; \cdot \; 10^{-2}$ & $(^{+   0.216}_{-  0.712})\cdot 10^{-2}$ & $(^{+   0.553}_{-  0.493})\cdot 10^{-2}$ & $(\pm   0.767)\cdot 10^{-2}$ & $(^{+   0.155}_{-  0.164})\cdot 10^{-2}$ &    1.05$\pm$    0.02 \\
 \hline
  80. --  120. &   92.06 &   1.657$\; \cdot \; 10^{-2}$ & $(^{+   0.072}_{-  0.182})\cdot 10^{-2}$ & $(^{+   0.128}_{-  0.091})\cdot 10^{-2}$ & $(\pm   0.170)\cdot 10^{-2}$ & $(^{+   0.040}_{-  0.050})\cdot 10^{-2}$ &    0.96$\pm$    0.02 \\
 \hline
 \end{tabular}
 \end{center}
 \caption{\label{table16} The QCD predictions for the differential dijet cross section as a
  function of $M_{jj}$. Other details are as described in the caption to
  Table~\ref{table13}.}
 \end{table}

 \begin{table} [!hb]
 \begin{center}
 \vspace*{0.7cm}
 \hspace*{-0.9cm}
 \begin{tabular}{|c|c|c|c|c|c|c|c|}
 \hline
 \hline
 $E_{TB}^{jet,1}$ range [GeV]&                                              
 $<E_{T,B}^{jet,1}>$ &  
 $d\sigma_{2+1}^{NLO}/dE_{T,B}^{jet,1}$~[pb/GeV] &                     
 $\Delta_{\mu_r}$ &
 $\Delta_{\alpha_s}$ &
 $\Delta^{\rm exp}_{pdf}$ &
 $\Delta^{\rm theo}_{pdf}$ &
 $C_{\rm had}$ $\pm$ $\Delta C_{\rm had}$ \\
 \hline
 \hline
   5. --    8. &    7.06 & \hspace*{-1.23cm}   3.722 & \hspace*{-1.23cm} $^{+   1.174}_{-  0.790}$ & \hspace*{-1.23cm} $^{+   0.429}_{-  0.397}$ & \hspace*{-1.23cm} $\pm   0.108$ & \hspace*{-1.23cm} $^{+   0.056}_{-  0.062}$ &    1.10$\pm$    0.01 \\
 \hline
   8. --   12. &    9.72 & \hspace*{-1.23cm}   6.637 & \hspace*{-1.23cm} $^{+   0.023}_{-  0.176}$ & \hspace*{-1.23cm} $^{+   0.304}_{-  0.366}$ & \hspace*{-1.23cm} $\pm   0.407$ & \hspace*{-1.23cm} $^{+   0.075}_{-  0.065}$ &    1.16$\pm$    0.03 \\
 \hline
  12. --   16. &   13.77 & \hspace*{-1.23cm}   2.905 & \hspace*{-1.23cm} $^{+   0.000}_{-  0.067}$ & \hspace*{-1.23cm} $^{+   0.145}_{-  0.163}$ & \hspace*{-1.23cm} $\pm   0.190$ & \hspace*{-1.23cm} $^{+   0.034}_{-  0.031}$ &    1.04$\pm$    0.00 \\
 \hline
  16. --   20. &   17.83 & \hspace*{-1.23cm}   1.420 & \hspace*{-1.23cm} $^{+   0.000}_{-  0.030}$ & \hspace*{-1.23cm} $^{+   0.079}_{-  0.083}$ & \hspace*{-1.23cm} $\pm   0.096$ & \hspace*{-1.23cm} $^{+   0.018}_{-  0.019}$ &    1.01$\pm$    0.01 \\
 \hline
  20. --   24. &   21.92 &   7.071$\; \cdot \; 10^{-1}$ & $(^{+   0.000}_{-  0.218})\cdot 10^{-1}$ & $(^{+   0.426}_{-  0.411})\cdot 10^{-1}$ & $(\pm   0.535)\cdot 10^{-1}$ & $(^{+   0.105}_{-  0.107})\cdot 10^{-1}$ &    1.00$\pm$    0.03 \\
 \hline
  24. --   32. &   27.28 &   2.951$\; \cdot \; 10^{-1}$ & $(^{+   0.001}_{-  0.118})\cdot 10^{-1}$ & $(^{+   0.199}_{-  0.175})\cdot 10^{-1}$ & $(\pm   0.214)\cdot 10^{-1}$ & $(^{+   0.051}_{-  0.056})\cdot 10^{-1}$ &    1.01$\pm$    0.03 \\
 \hline
  32. --   50. &   37.77 &   5.730$\; \cdot \; 10^{-2}$ & $(^{+   0.098}_{-  0.381})\cdot 10^{-2}$ & $(^{+   0.464}_{-  0.339})\cdot 10^{-2}$ & $(\pm   0.482)\cdot 10^{-2}$ & $(^{+   0.128}_{-  0.173})\cdot 10^{-2}$ &    1.01$\pm$    0.03 \\
 \hline
  50. --   80. &   57.65 &   3.908$\; \cdot \; 10^{-3}$ & $(^{+   0.280}_{-  0.609})\cdot 10^{-3}$ & $(^{+   0.324}_{-  0.162})\cdot 10^{-3}$ & $(\pm   0.441)\cdot 10^{-3}$ & $(^{+   0.094}_{-  0.244})\cdot 10^{-3}$ &    1.04$\pm$    0.03 \\
 \hline
 \end{tabular}
 \end{center}
 \vspace*{-0.2cm}
 \caption{\label{table17} The QCD predictions for the differential dijet cross section as a
  function of $E_{TB}^{jet,1}$. Other details are as described in the caption to
  Table~\ref{table13}.}
 \end{table}

 \begin{table} [!hb]
 \begin{center}
 \begin{tabular}{|c|c|c|c|c|c|c|c|}
 \hline
 \hline
 $\eta_B^{jet,1}$ range &                                              
 $<\eta_B^{jet,1}>$ &                                                  
 $d\sigma_{2+1}^{NLO}/d\eta_B^{jet,1}$~[pb] &                          
 $\Delta_{\mu_r}$ &
 $\Delta_{\alpha_s}$ &
 $\Delta^{\rm exp}_{pdf}$ &
 $\Delta^{\rm theo}_{pdf}$ &
 $C_{\rm had}$ $\pm$ $\Delta C_{\rm had}$ \\
 \hline
 \hline
-1.0 -- -0.5 & -0.68 & \hspace*{-1.00cm}   5.156 & \hspace*{-1.00cm} $^{+   0.305}_{-  0.294}$ & \hspace*{-1.00cm} $^{+   0.335}_{-  0.316}$ & \hspace*{-1.00cm} $\pm   0.257$ & \hspace*{-1.00cm} $^{+   0.054}_{-  0.046}$ &    1.47$\pm$    0.05 \\
 \hline
-0.5 --  0.0 & -0.21 &   1.972$\; \cdot \; 10^{ 1}$ & $(^{+   0.066}_{-  0.086})\cdot 10^{ 1}$ & $(^{+   0.115}_{-  0.120})\cdot 10^{ 1}$ & $(\pm   0.109)\cdot 10^{ 1}$ & $(^{+   0.024}_{-  0.022})\cdot 10^{ 1}$ &    1.34$\pm$    0.05 \\
 \hline
 0.0 --  0.5 &  0.26 &   3.596$\; \cdot \; 10^{ 1}$ & $(^{+   0.126}_{-  0.163})\cdot 10^{ 1}$ & $(^{+   0.205}_{-  0.222})\cdot 10^{ 1}$ & $(\pm   0.197)\cdot 10^{ 1}$ & $(^{+   0.046}_{-  0.044})\cdot 10^{ 1}$ &    1.11$\pm$    0.01 \\
 \hline
 0.5 --  1.0 &  0.73 &   3.136$\; \cdot \; 10^{ 1}$ & $(^{+   0.152}_{-  0.168})\cdot 10^{ 1}$ & $(^{+   0.194}_{-  0.205})\cdot 10^{ 1}$ & $(\pm   0.171)\cdot 10^{ 1}$ & $(^{+   0.041}_{-  0.041})\cdot 10^{ 1}$ &    1.02$\pm$    0.02 \\
 \hline
 1.0 --  1.5 &  1.22 &   1.888$\; \cdot \; 10^{ 1}$ & $(^{+   0.144}_{-  0.132})\cdot 10^{ 1}$ & $(^{+   0.129}_{-  0.134})\cdot 10^{ 1}$ & $(\pm   0.099)\cdot 10^{ 1}$ & $(^{+   0.025}_{-  0.026})\cdot 10^{ 1}$ &    0.99$\pm$    0.04 \\
 \hline
 1.5 --  2.0 &  1.71 & \hspace*{-1.00cm}   8.064 & \hspace*{-1.00cm} $^{+   0.884}_{-  0.724}$ & \hspace*{-1.00cm} $^{+   0.596}_{-  0.617}$ & \hspace*{-1.00cm} $\pm   0.413$ & \hspace*{-1.00cm} $^{+   0.106}_{-  0.112}$ &    1.00$\pm$    0.01 \\
 \hline
 2.0 --  2.5 &  2.18 & \hspace*{-1.00cm}   2.406 & \hspace*{-1.00cm} $^{+   0.384}_{-  0.287}$ & \hspace*{-1.00cm} $^{+   0.197}_{-  0.204}$ & \hspace*{-1.00cm} $\pm   0.120$ & \hspace*{-1.00cm} $^{+   0.031}_{-  0.033}$ &    1.08$\pm$    0.01 \\
 \hline
 \end{tabular}
 \end{center}
 \caption{\label{table18} The QCD predictions for the differential dijet cross section as a
  function of $\eta_B^{jet,1}$. Other details are as described in the caption to
  Table~\ref{table13}.}
 \end{table}

 \begin{table} [!hb]
 \begin{center}
 \vspace*{0.7cm}
 \hspace*{-0.9cm}
 \begin{tabular}{|c|c|c|c|c|c|c|c|}
 \hline
 \hline
 $E_{TB}^{jet,2}$ range [GeV]&                                              
 $<E_{T,B}^{jet,2}>$ &                                                 
 $d\sigma_{2+1}^{NLO}/dE_{T,B}^{jet,2}$~[pb/GeV] &                     
 $\Delta_{\mu_r}$ &
 $\Delta_{\alpha_s}$ &
 $\Delta^{\rm exp}_{pdf}$ &
 $\Delta^{\rm theo}_{pdf}$ &
 $C_{\rm had}$ $\pm$ $\Delta C_{\rm had}$ \\
 \hline
 \hline
   5. --    8. &    7.10 & \hspace*{-1.23cm}   3.623 & \hspace*{-1.23cm} $^{+   1.148}_{-  0.771}$ & \hspace*{-1.23cm} $^{+   0.425}_{-  0.390}$ & \hspace*{-1.23cm} $\pm   0.109$ & \hspace*{-1.23cm} $^{+   0.056}_{-  0.063}$ &    1.20$\pm$    0.02 \\
 \hline
   8. --   12. &    9.69 & \hspace*{-1.23cm}   6.737 & \hspace*{-1.23cm} $^{+   0.053}_{-  0.196}$ & \hspace*{-1.23cm} $^{+   0.315}_{-  0.376}$ & \hspace*{-1.23cm} $\pm   0.398$ & \hspace*{-1.23cm} $^{+   0.076}_{-  0.066}$ &    1.13$\pm$    0.02 \\
 \hline
  12. --   16. &   13.75 & \hspace*{-1.23cm}   2.877 & \hspace*{-1.23cm} $^{+   0.000}_{-  0.062}$ & \hspace*{-1.23cm} $^{+   0.139}_{-  0.159}$ & \hspace*{-1.23cm} $\pm   0.185$ & \hspace*{-1.23cm} $^{+   0.033}_{-  0.031}$ &    1.04$\pm$    0.01 \\
 \hline
  16. --   20. &   17.70 & \hspace*{-1.23cm}   1.367 & \hspace*{-1.23cm} $^{+   0.000}_{-  0.033}$ & \hspace*{-1.23cm} $^{+   0.071}_{-  0.076}$ & \hspace*{-1.23cm} $\pm   0.094$ & \hspace*{-1.23cm} $^{+   0.017}_{-  0.016}$ &    1.02$\pm$    0.02 \\
 \hline
  20. --   24. &   21.70 &   7.527$\; \cdot \; 10^{-1}$ & $(^{+   0.000}_{-  0.154})\cdot 10^{-1}$ & $(^{+   0.464}_{-  0.455})\cdot 10^{-1}$ & $(\pm   0.527)\cdot 10^{-1}$ & $(^{+   0.106}_{-  0.108})\cdot 10^{-1}$ &    1.00$\pm$    0.03 \\
 \hline
  24. --   32. &   27.18 &   3.020$\; \cdot \; 10^{-1}$ & $(^{+   0.000}_{-  0.098})\cdot 10^{-1}$ & $(^{+   0.210}_{-  0.184})\cdot 10^{-1}$ & $(\pm   0.212)\cdot 10^{-1}$ & $(^{+   0.052}_{-  0.057})\cdot 10^{-1}$ &    0.99$\pm$    0.03 \\
 \hline
  32. --   50. &   37.71 &   5.643$\; \cdot \; 10^{-2}$ & $(^{+   0.117}_{-  0.410})\cdot 10^{-2}$ & $(^{+   0.446}_{-  0.327})\cdot 10^{-2}$ & $(\pm   0.476)\cdot 10^{-2}$ & $(^{+   0.125}_{-  0.165})\cdot 10^{-2}$ &    1.02$\pm$    0.03 \\
 \hline
  50. --   80. &   57.61 &   3.867$\; \cdot \; 10^{-3}$ & $(^{+   0.288}_{-  0.618})\cdot 10^{-3}$ & $(^{+   0.315}_{-  0.161})\cdot 10^{-3}$ & $(\pm   0.440)\cdot 10^{-3}$ & $(^{+   0.093}_{-  0.225})\cdot 10^{-3}$ &    0.97$\pm$    0.03 \\
 \hline
 \end{tabular}
 \end{center}
 \vspace*{-0.2cm}
 \caption{\label{table19} The QCD predictions for the differential dijet cross section as a
  function of $E_{TB}^{jet,2}$. Other details are as described in the caption to
  Table~\ref{table13}.}
 \end{table}

 \begin{table} [!hb]
 \begin{center}
 \begin{tabular}{|c|c|c|c|c|c|c|c|}
 \hline
 \hline
 $\eta_B^{jet,2}$ range &                                              
 $<\eta_B^{jet,2}>$ &                                                  
 $d\sigma_{2+1}^{NLO}/d\eta_B^{jet,2}$~[pb] &                          
 $\Delta_{\mu_r}$ &
 $\Delta_{\alpha_s}$ &
 $\Delta^{\rm exp}_{pdf}$ &
 $\Delta^{\rm theo}_{pdf}$ &
 $C_{\rm had}$ $\pm$ $\Delta C_{\rm had}$ \\
 \hline
 \hline
-2.0 -- -1.5 & -1.69 & \hspace*{-1.00cm}   4.973 & \hspace*{-1.00cm} $^{+   0.354}_{-  0.317}$ & \hspace*{-1.00cm} $^{+   0.343}_{-  0.309}$ & \hspace*{-1.00cm} $\pm   0.264$ & \hspace*{-1.00cm} $^{+   0.054}_{-  0.051}$ &    1.51$\pm$    0.05 \\
 \hline
-1.5 -- -1.0 & -1.20 &   1.972$\; \cdot \; 10^{ 1}$ & $(^{+   0.151}_{-  0.135})\cdot 10^{ 1}$ & $(^{+   0.135}_{-  0.133})\cdot 10^{ 1}$ & $(\pm   0.104)\cdot 10^{ 1}$ & $(^{+   0.024}_{-  0.022})\cdot 10^{ 1}$ &    1.40$\pm$    0.00 \\
 \hline
-1.0 -- -0.5 & -0.74 &   4.169$\; \cdot \; 10^{ 1}$ & $(^{+   0.154}_{-  0.194})\cdot 10^{ 1}$ & $(^{+   0.234}_{-  0.258})\cdot 10^{ 1}$ & $(\pm   0.238)\cdot 10^{ 1}$ & $(^{+   0.051}_{-  0.048})\cdot 10^{ 1}$ &    1.21$\pm$    0.03 \\
 \hline
-0.5 --  0.0 & -0.28 &   3.427$\; \cdot \; 10^{ 1}$ & $(^{+   0.132}_{-  0.164})\cdot 10^{ 1}$ & $(^{+   0.195}_{-  0.216})\cdot 10^{ 1}$ & $(\pm   0.180)\cdot 10^{ 1}$ & $(^{+   0.044}_{-  0.043})\cdot 10^{ 1}$ &    0.99$\pm$    0.02 \\
 \hline
 0.0 --  0.5 &  0.20 &   1.558$\; \cdot \; 10^{ 1}$ & $(^{+   0.100}_{-  0.099})\cdot 10^{ 1}$ & $(^{+   0.109}_{-  0.110})\cdot 10^{ 1}$ & $(\pm   0.077)\cdot 10^{ 1}$ & $(^{+   0.023}_{-  0.024})\cdot 10^{ 1}$ &    0.92$\pm$    0.01 \\
 \hline
 0.5 --  1.0 &  0.69 & \hspace*{-1.00cm}   4.185 & \hspace*{-1.00cm} $^{+   0.459}_{-  0.377}$ & \hspace*{-1.00cm} $^{+   0.371}_{-  0.342}$ & \hspace*{-1.00cm} $\pm   0.209$ & \hspace*{-1.00cm} $^{+   0.071}_{-  0.079}$ &    0.96$\pm$    0.00 \\
 \hline
 1.0 --  1.5 &  1.19 &   9.857$\; \cdot \; 10^{-1}$ & $(^{+   2.051}_{-  1.459})\cdot 10^{-1}$ & $(^{+   1.041}_{-  0.945})\cdot 10^{-1}$ & $(\pm   0.454)\cdot 10^{-1}$ & $(^{+   0.170}_{-  0.201})\cdot 10^{-1}$ &    1.36$\pm$    0.03 \\
 \hline
 \end{tabular}
 \end{center}
 \caption{\label{table20} The QCD predictions for the differential dijet cross section as a
  function of $\eta_B^{jet,2}$. Other details are as described in the caption to
  Table~\ref{table13}.}
 \end{table}

 \begin{table} [!hb]
 \begin{center}
 \begin{tabular}{|c|c|c|c|c|c|c|}
 \hline
 \hline
 $Q^2$ range [GeV$^2$]&                                                         
 $<Q^2>$ &                                                             
 $d\sigma_{tot}^{NLO}/dQ^2$~[pb/GeV$^2$] &                             
 $\Delta_{\mu_r}$ &
 $\Delta_{\alpha_s}$ &
 $\Delta^{\rm exp}_{pdf}$ &
 $\Delta^{\rm theo}_{pdf}$ \\
 \hline
 \hline
     470. --      800. &      599. & \hspace*{-1.20cm}   1.180 & \hspace*{-1.20cm} $^{+   0.008}_{-  0.011}$ & \hspace*{-1.20cm} $^{+   0.002}_{-  0.006}$ & \hspace*{-1.20cm} $\pm   0.032$ & \hspace*{-1.20cm} $^{+   0.017}_{-  0.014}$ \\
 \hline
     800. --     1500. &     1059. &   2.756$\; \cdot \; 10^{-1}$ & $(^{+   0.016}_{-  0.021})\cdot 10^{-1}$ & $(^{+   0.006}_{-  0.012})\cdot 10^{-1}$ & $(\pm   0.067)\cdot 10^{-1}$ & $(^{+   0.034}_{-  0.026})\cdot 10^{-1}$ \\ 
 \hline
    1500. --     2500. &     1890. &   6.282$\; \cdot \; 10^{-2}$ & $(^{+   0.028}_{-  0.036})\cdot 10^{-2}$ & $(^{+   0.014}_{-  0.015})\cdot 10^{-2}$ & $(\pm   0.146)\cdot 10^{-2}$ & $(^{+   0.064}_{-  0.046})\cdot 10^{-2}$ \\ 
 \hline
    2500. --     5000. &     3363. &   1.257$\; \cdot \; 10^{-2}$ & $(^{+   0.004}_{-  0.005})\cdot 10^{-2}$ & $(^{+   0.001}_{-  0.000})\cdot 10^{-2}$ & $(\pm   0.030)\cdot 10^{-2}$ & $(^{+   0.009}_{-  0.007})\cdot 10^{-2}$ \\ 
 \hline
    5000. --    20000. &     7801. &   7.003$\; \cdot \; 10^{-4}$ & $(^{+   0.010}_{-  0.012})\cdot 10^{-4}$ & $(^{+   0.057}_{-  0.035})\cdot 10^{-4}$ & $(\pm   0.188)\cdot 10^{-4}$ & $(^{+   0.003}_{-  0.031})\cdot 10^{-4}$ \\ 
 \hline
 \end{tabular}
 \end{center}
 \caption{\label{table21} The QCD predictions for the differential inclusive cross section as a
  function of $Q^2$. Other details are as described in the caption to
  Table~\ref{table13}.}
 \end{table}

 \begin{table} [!hb]
 \begin{center}
 \begin{tabular}{|c|c|c|c|c|c|c|c|}
 \hline
 \hline
 $Q^2$ range [GeV$^2$]&                                                         
 $<Q^2>$ &                                                             
 $d\sigma_{2+1}^{NLO}/dQ^2$~[pb/GeV$^2$] &                             
 $\Delta_{\mu_r}$ &
 $\Delta_{\alpha_s}$ &
 $\Delta^{\rm exp}_{pdf}$ &
 $\Delta^{\rm theo}_{pdf}$ &
 $C_{\rm had}$ $\pm$ $\Delta C_{\rm had}$ \\
 \hline
 \hline
     470. --      800. &      604. &   8.985$\; \cdot \; 10^{-2}$ & $(^{+   0.583}_{-  0.577})\cdot 10^{-2}$ & $(^{+   0.539}_{-  0.606})\cdot 10^{-2}$ & $(\pm   0.388)\cdot 10^{-2}$ & $(^{+   0.118}_{-  0.121})\cdot 10^{-2}$ &    1.11$\pm$    0.01 \\
 \hline
     800. --     1500. &     1071. &   2.597$\; \cdot \; 10^{-2}$ & $(^{+   0.128}_{-  0.139})\cdot 10^{-2}$ & $(^{+   0.167}_{-  0.172})\cdot 10^{-2}$ & $(\pm   0.098)\cdot 10^{-2}$ & $(^{+   0.034}_{-  0.031})\cdot 10^{-2}$ &    1.09$\pm$    0.01 \\
 \hline
    1500. --     2500. &     1901. &   7.208$\; \cdot \; 10^{-3}$ & $(^{+   0.270}_{-  0.324})\cdot 10^{-3}$ & $(^{+   0.484}_{-  0.460})\cdot 10^{-3}$ & $(\pm   0.263)\cdot 10^{-3}$ & $(^{+   0.095}_{-  0.080})\cdot 10^{-3}$ &    1.07$\pm$    0.01 \\
 \hline
    2500. --     5000. &     3420. &   1.734$\; \cdot \; 10^{-3}$ & $(^{+   0.051}_{-  0.067})\cdot 10^{-3}$ & $(^{+   0.114}_{-  0.101})\cdot 10^{-3}$ & $(\pm   0.070)\cdot 10^{-3}$ & $(^{+   0.022}_{-  0.019})\cdot 10^{-3}$ &    1.05$\pm$    0.01 \\
 \hline
    5000. --    20000. &     8095. &   1.281$\; \cdot \; 10^{-4}$ & $(^{+   0.038}_{-  0.046})\cdot 10^{-4}$ & $(^{+   0.068}_{-  0.057})\cdot 10^{-4}$ & $(\pm   0.055)\cdot 10^{-4}$ & $(^{+   0.008}_{-  0.014})\cdot 10^{-4}$ &    1.04$\pm$    0.01 \\
 \hline
 \end{tabular}
 \end{center}
 \caption{\label{table22} The QCD predictions for the differential dijet cross section as a
  function of $Q^2$. Other details are as described in the caption to
  Table~\ref{table13}.}
 \end{table}

 \begin{table} [!hb]
 \begin{center}
 \begin{tabular}{|c|c|c|c|c|c|c|c|}
 \hline
 \hline
 $Q^2$ range [GeV$^2$]&                                                         
 $<Q^2>$ &                                                             
 $R_{2+1}$ &                                                           
 $\Delta_{\mu_r}$ &
 $\Delta_{\alpha_s}$ &
 $\Delta^{\rm exp}_{pdf}$ &
 $\Delta^{\rm theo}_{pdf}$ &
 $C_{\rm had}$ $\pm$ $\Delta C_{\rm had}$ \\
 \hline
 \hline
     470. --      800. &      604. &   7.615$\; \cdot \; 10^{-2}$ & $(^{+   0.569}_{-  0.538})\cdot 10^{-2}$ & $(^{+   0.443}_{-  0.475})\cdot 10^{-2}$ & $(\pm   0.122)\cdot 10^{-2}$ & $(^{+   0.000}_{-  0.017})\cdot 10^{-2}$ &    1.11$\pm$    0.01 \\
 \hline
     800. --     1500. &     1071. &   9.424$\; \cdot \; 10^{-2}$ & $(^{+   0.539}_{-  0.556})\cdot 10^{-2}$ & $(^{+   0.583}_{-  0.586})\cdot 10^{-2}$ & $(\pm   0.122)\cdot 10^{-2}$ & $(^{+   0.007}_{-  0.025})\cdot 10^{-2}$ &    1.09$\pm$    0.01 \\
 \hline
    1500. --     2500. &     1901. &   1.148$\; \cdot \; 10^{-1}$ & $(^{+   0.050}_{-  0.056})\cdot 10^{-1}$ & $(^{+   0.074}_{-  0.071})\cdot 10^{-1}$ & $(\pm   0.015)\cdot 10^{-1}$ & $(^{+   0.003}_{-  0.004})\cdot 10^{-1}$ &    1.07$\pm$    0.01 \\
 \hline
    2500. --     5000. &     3420. &   1.380$\; \cdot \; 10^{-1}$ & $(^{+   0.047}_{-  0.058})\cdot 10^{-1}$ & $(^{+   0.089}_{-  0.081})\cdot 10^{-1}$ & $(\pm   0.022)\cdot 10^{-1}$ & $(^{+   0.007}_{-  0.007})\cdot 10^{-1}$ &    1.05$\pm$    0.01 \\
 \hline
    5000. --    20000. &     8095. &   1.829$\; \cdot \; 10^{-1}$ & $(^{+   0.057}_{-  0.069})\cdot 10^{-1}$ & $(^{+   0.106}_{-  0.096})\cdot 10^{-1}$ & $(\pm   0.029)\cdot 10^{-1}$ & $(^{+   0.011}_{-  0.012})\cdot 10^{-1}$ &    1.04$\pm$    0.01 \\
 \hline
 \end{tabular}
 \end{center}
 \caption{\label{table23} The QCD predictions for the dijet fraction as a
  function of $Q^2$. Other details are as described in the caption to
  Table~\ref{table13}.}
 \end{table}
\end{landscape}

\catcode`\@=11 
\renewcommand{\@makecaption}[2]%
  {\def\baselinestretch{0.95}%
   \vspace{10.pt}
   \setlength{\@captionwidth}{\localtextwidth}
   \addtolength{\@captionwidth}{-\@captionmargin}
   \sbox{\tmpbox}{{\bf #1:}{\rm #2}}%
   \ifthenelse{\lengthtest{\wd\tmpbox > \@captionwidth}}%
   {\centerline{\parbox[t]{\@captionwidth}%
   {\tolerance=2000\normalsize%
    {\bf #1:}\hspace{\@captionitemtextsep}{\rm #2}}}}%
   {\centerline{{\bf #1:}\kern1.em{\rm #2}}}}
\catcode`\@=12 

\newpage
\clearpage
\begin{table}
\begin{center}
\begin{tabular}{|c|c|c|c|c|c|}
 \hline
 \hline
 $\langle Q \rangle$ {\rm [GeV]} &
 $\alpha_s(\langle Q \rangle)$ &
 $\Delta\alpha_s^{\rm stat}$ &
 $\Delta\alpha_s^{\rm syst}$ &
 $\Delta\alpha_s^{\rm ES}$ &
 $\Delta \alpha_s^{\rm Th}$ \\ 
 \hline
 \hline
 24.6 & 0.1436 & 0.0047 & $^{+0.0032}_{-0.0066}$ & $^{+0.0045}_{-0.0041}$ & 
$^{+0.0105}_{-0.0079}$ \\
 \hline
 32.7 & 0.1396 & 0.0048 & $^{+0.0019}_{-0.0036}$ & $^{+0.0034}_{-0.0034}$ & 
$^{+0.0078}_{-0.0059}$ \\
 \hline
 43.6 & 0.1306 & 0.0063 & $^{+0.0041}_{-0.0044}$ & $^{+0.0023}_{-0.0024}$ & 
$^{+0.0054}_{-0.0040}$ \\
 \hline
 58.5 & 0.1276 & 0.0075 & $^{+0.0027}_{-0.0041}$ & $^{+0.0019}_{-0.0016}$ & 
$^{+0.0046}_{-0.0033}$ \\
 \hline
 90.0 & 0.1149 & 0.0111 & $^{+0.0056}_{-0.0099}$ & $^{+0.0009}_{-0.0013}$ & 
$^{+0.0037}_{-0.0028}$ \\
 \hline
\end{tabular}
\end{center}
\caption{\label{table24} The $\alpha_s$ values as determined from the QCD fit 
to the measured dijet fraction $R_{2+1}$ as a function of $Q$. For each bin in
$Q^2$, the mean value $\langle Q \rangle$, the extracted value of the
strong coupling constant, $\alpha_s(\langle Q \rangle)$, the statistical 
uncertainty, $\Delta\alpha_s^{\rm stat}$, the systematic uncertainty
(not) associated with the energy scale of the jets, $\Delta\alpha_s^{\rm ES}$
($\Delta\alpha_s^{\rm syst}$), and the total theoretical uncertainty,
$\Delta \alpha_s^{\rm Th}$, are given.}
\end{table}

%
%
\newpage
\clearpage
\parskip 0mm

\newpage
\clearpage
\parskip 0mm

\newpage
\clearpage
\parskip 0mm

\newpage
\clearpage
\parskip 0mm

\newpage
\clearpage
\parskip 0mm


\begin{thebibliography}{99}
\bibitem{oldalp}
 ZEUS Collaboration, M. Derrick \etal , \Journal{\PLB}{363}{1995}{201}.

\bibitem{factor}
 See e.g. J.C. Collins, D.E. Soper and G. Sterman, in {\it Perturbative Quantum
 Chromodynamics}, edited by A.H. Mueller, World Scientific, Singapore (1989),
 p. 1 and references therein.

\bibitem{dglap}
 V.N. Gribov and L.N. Lipatov, \Journal{\SJN}{15}{1972}{438}; \\
 L.N. Lipatov, \Journal{\SJN}{20}{1975}{94}; \\
 Y.L. Dokshitzer, \Journal{\SPJ}{46} {1977} {641}; \\
 G. Altarelli and G. Parisi, \Journal{\NPB}{126}{1977}{298}.

\bibitem{singu}
 S.D. Ellis, Z. Kunszt and D.E. Soper, \Journal{\PRD}{40}{1989}{2188}
 and \Journal{\PRL}{69}{1992}{1496};\\
 Z. Kunszt and D.E. Soper, \Journal{\PRD}{46}{1992}{192};\\
 W.T. Giele and E.W.N. Glover, \Journal{\PRD}{46}{1992}{1980};\\
 W.T. Giele, E.W.N. Glover and D.A. Kosower, \Journal{\NPB}{403}{1993}{633};\\
 S. Frixione, Z. Kunszt and A. Signer, \Journal{\NPB}{467}{1996}{399};\\
 Z. Nagy and Z. Trocsanyi, \Journal{\NPB}{486}{1997}{189}.

\bibitem{disent}
 S. Catani and M. H. Seymour, \Journal{\NPB}{485}{1997}{291}
 and Erratum B~510 (1997) 503.

\bibitem{mepjet}
 E. Mirkes and D. Zeppenfeld, \Journal{\PLB}{380}{1996}{205}; \\
 E. Mirkes, TTP-97-39 (hep-ph/9711224).

\bibitem{disaster}
 D. Graudenz, hep-ph/9710244.

\bibitem{jetvip}
 B. P\"otter, \Journal{\CPC}{119}{1999}{45}.

\bibitem{status}
 ZEUS Collaboration, U. Holm (ed.), {\it The ZEUS Detector}, Status Report
 (unpublished), DESY, 1993, available on WWW: \texttt{http://www-zeus.desy.de/bluebook/bluebook.html}.

\bibitem{zeuscal}
 M. Derrick \etal , \Journal{\NIM}{309}{1991}{77}; \\ 
 A. Andresen \etal , \Journal{\NIM}{309}{1991}{101}; \\ 
 A. Caldwell \etal , \Journal{\NIM}{321}{1992}{356}; \\ 
 A. Bernstein \etal , \Journal{\NIM}{336}{1993}{23}.

\bibitem{zeusctd}
 N. Harnew \etal , \Journal{\NIM}{279}{1989}{290}; \\ 
 B. Foster \etal , Nucl. Phys. B (Proc. Suppl.) {32} (1993) 181; \\ 
 B. Foster \etal , \Journal{\NIM}{338}{1994}{254}.

\bibitem{zeuslumi}
 J. Andruszk\'ow \etal , DESY 92-066 (1992); \\ 
 ZEUS Collaboration, M. Derrick \etal , \Journal{\ZPC}{63}{1994}{391}.

\bibitem{heracles}
 K. Kwiatkowski, H. Spiesberger and H.-J. M\"ohring,
 \Journal{\CPC}{69}{1992}{155}.

\bibitem{django}
 K. Charchu{\l}a, G.A. Schuler, and H. Spiesberger,
 \Journal{\CPC}{81}{1994}{381}.

\bibitem{cdm}
 Y. Azimov \etal , \Journal{\PLB}{165}{1985}{147}; \\
 G. Gustafson, \Journal{\PLB}{175}{1986}{453}; \\ 
 G. Gustafson and U. Petersson, \Journal{\NPB}{306}{1988}{746}; \\ 
 B. Andersson, G. Gustafson and L. L\"onnblad, \Journal{\ZPC}{43}{1989}{625}.

\bibitem{modlon}
 L. L\"onnblad, in {\it Proc. Workshop on Monte Carlo Generators for HERA Physics,
 Apr. 1998}, eds.  A.T. Doyle, G. Grindhammer, G. Ingelman and H. Jung,
 p. 47, Hamburg, Germany, DESY (1999) and hep-ph/9908368.

\bibitem{ariadne}
 L. L\"onnblad, \Journal{\CPC}{71}{1992}{15};
 \Journal{\ZPC}{65}{1995}{285}.

\bibitem{lepto}
 G. Ingelman, A. Edin and J. Rathsman, \Journal{\CPC}{101}{1997}{108}.

\bibitem{lund}
 B. Andersson \etal , \Journal{\PRE}{97}{1983}{31}.

\bibitem{pythia}
 H.-U. Bengtsson and T. Sj\"ostrand,
 \Journal{\CPC}{46}{1987}{43}; \\ 
 T. Sj\"ostrand, \Journal{\CPC}{82}{1994}{74}.

\bibitem{herwig}
 G. Marchesini \etal , \Journal{\CPC}{67}{1992}{465}. 

\bibitem{cluster}
 B.R. Webber, \Journal{\NPB}{238}{1984}{492}.

\bibitem{geant}
 R. Brun \etal , CERN-DD/EE/84-1 (1987).

\bibitem{terrano}
 R.K. Ellis, D.A. Ross and A.E. Terrano, \Journal{\NPB}{178}{1981}{421}.

\bibitem{alpworld}
 Particle Data Group, D.E. Groom \etal , \Journal{\EPC}{15}{2000}{1}.

\bibitem{botje}
 M. Botje, \Journal{\EPC}{14}{2000}{285}.

\bibitem{zenc99}
 ZEUS Collaboration, J. Breitweg \etal , \Journal{\EPC}{11}{1999}{427}.

\bibitem{dameth}
 S. Bentvelsen, J. Engelen, and P. Kooijman, in {\it Proc. Workshop on Physics at 
 HERA, Oct. 1991}, Volume 1, W. Buchm\"uller and G. Ingelman (eds.),
 DESY (1992), p. 23; \\ K. C. Hoeger, {\it ibid.}, p. 43.

\bibitem{kt}
 S. Catani \etal, \Journal{\NPB}{406}{1993}{187}; \\ 
 S.D. Ellis and D.E. Soper, \Journal{\PRD}{48}{1993}{3160}.

\bibitem{breit}
 R.P. Feynman, {\it Photon-Hadron Interactions}, Benjamin, (1972); \\ 
 K.H. Streng, T.F. Walsh and P.M. Zerwas, \Journal{\ZPC}{2}{1979}{237}.

\bibitem{snow}
 J. Huth \etal , in {\it Proc. of the 1990 DPF Summer Study on High Energy
 Physics}, ed. E.L. Berger, p. 134, Snowmass, Colorado, World
Scientific (1992).

\bibitem{zeusjs}
 ZEUS Collaboration, J. Breitweg \etal , \Journal{\EPC}{8}{1999}{367}.

\bibitem{etthesis}
 E. Tassi, Ph.D. Thesis, University of Hamburg (2001), in preparation.

\bibitem{klasen}
 M. Klasen and G. Kramer, \Journal{\PLB}{366}{1996}{385}; \\ 
 S. Frixione and G. Ridolfi, \Journal{\NPB}{507}{1997}{315};\\
 B. P\"otter, \Journal{\CPC}{133}{2000}{105}.

\bibitem{cteq4}
 H.L. Lai \etal ,  \Journal{\PRD}{55}{1997}{1280}.

\bibitem{mrst}
 A.D. Martin \etal ,  \Journal{\EPC}{4}{1998}{463}.

\bibitem{bethke}
 S. Bethke,  \Journal{\JPG}{G26}{2000}{R27}.

\bibitem{h1alp}
 H1 Collaboration, C. Adloff \etal , DESY 00-145 (2000) and DESY 00-181 (2000).
\end{thebibliography}
\end{document}